\documentclass[iop]{emulateapj}
\usepackage{natbib}
\usepackage{amsmath}
\usepackage{float}

\newcommand{\kms}{km s$^{-1}$}

\slugcomment{Author version of ApJ, 779, 121}

\shorttitle{MUSCLE W49. Data and Mass Structure.}
\shortauthors{Galv\'an-Madrid et al.}

\begin{document}


\title
{
MUSCLE W49 : A Multi-Scale Continuum and Line Exploration of the Most 
Luminous Star Formation Region in the Milky Way. I. 
Data and The Mass Structure of the Giant Molecular Cloud.  
}






\author{
R. Galv\'an-Madrid\altaffilmark{1}, 	
H. B. Liu\altaffilmark{2},
Z.-Y. Zhang\altaffilmark{3,4},
J. E. Pineda\altaffilmark{1,5,6}, 
T.-C. Peng\altaffilmark{1},
Q. Zhang\altaffilmark{7},  
E. R. Keto\altaffilmark{7},
P. T. P. Ho\altaffilmark{2,7},  
L. F. Rodr\'iguez\altaffilmark{8,9}, 
L. Zapata\altaffilmark{8}, 
T. Peters\altaffilmark{10}, 
C. G. De Pree\altaffilmark{11}
}


\altaffiltext{1}{European Southern Observatory, Karl-Schwarzschild-Str. 2, 85748 Garching, Germany.}
\altaffiltext{2}{Academia Sinica Institute of Astronomy and Astrophysics, P.O. Box 23-141, 
Taipei 106.}
\altaffiltext{3}{Max-Planck Institut f\"ur Radioastronomie, Auf dem H\"ugel 69, 53121, Bonn, 
Germany.}
\altaffiltext{4}{Purple Mountain Observatory, CAS, 2 West Beijing Road, Nanjing 210008, China.}
\altaffiltext{5}{UK ARC Node, Jodrell Bank Centre for Astrophysics, School of Physics and Astronomy, University of Manchester, Manchester, 
M13 9PL, UK.}
\altaffiltext{6}{Institute for Astronomy, ETH Zurich, Wolfgang-Pauli-Strasse 27, CH-8093 Zurich, Switzerland.}
\altaffiltext{7}{Harvard-Smithsonian Center for Astrophysics, 60 Garden Street, Cambridge, 
MA 02138, USA.}
\altaffiltext{8}{Centro de Radioastronom\'ia y Astrof\'isica, UNAM, A.P. 3-72 Xangari, Morelia 58089, Mexico.}
\altaffiltext{9}{Astronomy Department, Faculty of Science, King Abdulaziz University, PO Box 80203, 21589, Jeddah, Saudi Arabia.}
\altaffiltext{10}{Institut f\"{u}r Theoretische Physik, Universit\"{a}t Z\"{u}rich,
Winterthurerstrasse 190, CH-8057 Z\"{u}rich, Switzerland.}
\altaffiltext{11}{Department of Physics and Astronomy, Agnes Scott College, Decatur, GA 30030, USA}


\begin{abstract}
The Multi-Scale Continuum and Line Exploration of W49 (MUSCLE W49) is a comprehensive gas 
and dust survey of the giant molecular cloud (GMC) of W49A, the most 
luminous star-formation region in the Milky Way. 
The project covers, for the first time, the entire GMC at different scales and angular resolutions. 
In this paper we present: 
1) an all-configuration SMA mosaic in the 230-GHz (1.3-mm) band covering the central 
$\sim 3\arcmin \times 3\arcmin$ ($\sim 10$ pc, known as W49N), where most of the embedded massive stars reside; and 
2) PMO 14m telescope observations in the 90-GHz band, covering the 
entire GMC with maps of up to $\sim 35\arcmin \times 35\arcmin$ in size, or $\sim 113$ pc. 
We also make use of archival data from the VLA, JCMT-SCUBA, IRAM 30m, and the CSO BOLOCAM Galactic Plane Survey. 
We derive the basic physical parameters of the GMC at all scales. Our main findings are as follows: 
1) The W49 GMC is one of the most massive in the Galaxy, with a total mass $M_\mathrm{gas}\sim1.1\times 10^6$ M$_\odot$ 
within a radius of 60 pc. Within a radius of 6 pc, the total gas mass is $M_\mathrm{gas}\sim2\times 10^5$ M$_\odot$. 
At these scales only $\sim 1\%$ of the material is photoionized. The mass reservoir is sufficient to form several 
young massive clusters (YMCs) as massive as a globular cluster.   
2) The mass of the GMC is distributed in a hierarchical network of filaments. 
At scales $<10$ pc, a triple, centrally condensed structure peaks toward the ring of HC HII regions in 
W49N. This structure extends to scales from $\sim 10$ to 100 pc through filaments that radially converge toward W49N 
and its less prominent neighbor W49S. The W49A starburst most likely formed 
from global gravitational contraction with localized collapse in a  "hub-filament'' geometry.  
3) Currently, feedback from the central YMCs (with a present mass $M_\mathrm{cl} \gtrsim 5\times10^4$ M$_\odot$) 
is still not enough to entirely disrupt the GMC, but further stellar mass growth could be enough to allow radiation 
pressure to clear the cloud and halt star formation. 
4) The resulting stellar content will probably remain as a gravitationally bound massive star cluster, or a small 
system of bound clusters.
\end{abstract}


\keywords{Hii regions – ISM: individual objects (W49A,W49N,W49S,W49SW) – stars: formation – stars: massive 
– open clusters and associations: general – ISM: clouds – galaxies: starburst – galaxies: star clusters}



\section{Introduction} \label{sec:intro}

Young massive clusters (YMCs) have stellar masses $M_\mathrm{cl}>10^4$ M$_\odot$, sizescales of a 
few pc,  and are 
younger than a few crossing times (a few $\times 100$ Myr). They probably represent 
the young end of the so-called super star clusters (SSCs) commonly found in starbursting galaxies 
through high-resolution imaging with the {\it Hubble Space Telescope} \citep[e.g.,][]{Whitmore93,Meurer95,Ho97}. 
It is also possible that some of them are young analogues of globular clusters (GCs), provided that they remain 
gravitationally bound for timescales comparable to the Hubble time. Recent reviews are those of \cite{Turner09} and 
\cite{PZ10}.

Understanding how YMCs 
form is not only an interesting problem on itself, but is key to understand the properties of star forming galaxies elsewhere. 
In recent years there have been several studies of deeply embedded YMCs in nearby galaxies that address this problem. 
Some studies have targeted the progenitor giant molecular clouds (GMCs) \citep[e.g.,][]{Keto05,Santangelo09,Wei12}. 
\cite{Johnson01} suggested that the radio continuum sources embedded in some of these 
extragalactic GMCs trace the massive star content, making them analogues to ensembles of ultra compact (UC) and 
hypercompact (HC) HII regions embedded in the GMCs that harbor the most luminous star formation regions in the 
Milky Way \citep[e.g.,][]{HH81,Kurtz94}. 
However, even in nearby galaxies within 10 Mpc, subarcsecond resolution studies have physical resolutions not better 
than $\sim 10$ pc.

An alternative approach is to look for embedded YMCs in our Galaxy, which can be well resolved, with the disadvantage 
that there are only a few of them \citep[e.g.,][]{Quang11,Liu12a,NL13,Fukui13}. 
Deeply embedded, very luminous ($L_\mathrm{bol}>10^7$ L$_\odot$) star formation regions 
stand out as the obvious candidates to be active YMC forming sites.
W49A at Galactic coordinates $l=43.1^\circ$, $b=0.0^\circ$ 
is the most luminous star formation region in the Milky Way \citep[$L\sim10^{7.2}$ L$_\odot$,][]{Sievers91}, 
embedded in one of the most massive 
giant molecular clouds (GMCs), $M_\mathrm{gas}\sim10^6$ M$_\odot$ \citep{Simon01,Miyawaki09}. 
The GMC has an extent of $l>100$ pc, 
but all the prominent 
star formation resides in the central $\sim 20$ pc. This inner region contains the well known massive star formation 
regions W49 north (W49N), 
W49 south 
(W49S, $\sim 2\arcmin$ southeast of W49N), and W49 southwest (W49SW, $\sim 1.5\arcmin$ southwest of W49N). 
The most prominent by far is W49N, 
hosting the well known ring of HC and UCHIIs \citep{Welch87,DePree97} within a radius of a few pc. 
Part of the stellar population in W49N is already visible in the near-IR and 
its mass 
has been estimated at $M_\mathrm{cl}\gtrsim4\times10^4$ M$_\odot$ \citep{HomeierAlves05}, 
whereas the part associated with the  most compact HII regions is not 
even visible in the mid-IR \citep{Smith09}. W49N also hosts the most luminous water maser in the Galaxy \citep{Gwinn92}. 
Those authors determined 
a direct parallax distance from Earth of $d=11.4\pm1.2$ kpc. 
This measurement has been recently improved to $11.1\substack{+0.9 \\ -0.8}$ \citep{Zhang13}. 
We use the latter value through this paper.  

Several possible ideas have been given to explain the prodigious star formation in W49A. 
\cite{Welch87} proposed that the double-peaked profile seen in most molecular lines toward the center of W49N is due to 
large-scale collapse toward the central ring of HC HII regions. 
On the other hand, \cite{Miyawaki86}, \cite{Serabyn93}, and \cite{BuckleyWT96} 
interpreted it as a cloud-cloud collision. \cite{Peng10} showed evidence for expanding shells in the center of W49N, and proposed 
that these are the triggering factor for star formation in the whole region. 

In this paper we introduce our Multiscale Continuum and Line Exploration of W49 (MUSCLE W49), a project aimed at 
mapping the W49 GMC 
from its full scale ($\gtrsim 100$ pc) down to the scales of individual star-forming cores  ($\lesssim 0.03$ pc). 
Multiple molecular and hydrogen recombination lines (RLs) are observed, as well as dust and free-free 
continuum.
When we refer to "W49" or the "W49 GMC" we mean the full W49 giant molecular cloud, 
when we use "W49A" we refer to the central part of 
the GMC that is actively forming YMCs. When necessary, we refer individually to the common labeling of the subcomponents of W49A: 
W49N, W49S, and W49SW. 
This paper presents the bulk of the data set and derives the mass structure of the cloud at all scales. 
Upcoming papers will deal with further scientific analysis:
a multi-scale analysis of the dynamics, a comparison of hot cores and hypercompact HII 
regions using the subarcsecond resolution mosaics of W49N, and  a quantitative comparison with extragalactic star formation. 
Section \ref{sec:obs} and Appendix \ref{App-A} describe the observations. Section \ref{sec:data} presents the data. In Section 
\ref{mass-GMC} we derive the mass structure of the GMC, and in Section \ref{sec:disruption} we discuss on the GMC and cluster disruption. 
We list our conclusions in Section \ref{sec:concl}. Appendices \ref{App-B} and \ref{App-C} show the rest of the line maps from the PMO and SMA 
observations, respectively. Appendix \ref{App-D} describes our method to obtain the gas surface density maps using CO and its isotopologues,
and to derive gas masses from the dust continuum. 

\section{Observations} \label{sec:obs}


\subsection{PMO 14m}

Mapping observations were made from March 2011 to July 2011 with the 14-m
telescope of Purple Mountain Observatory\footnote{http://www.dlh.pmo.cas.cn/} 
(hereafter PMO 14m), located at Delingha, Qinghai, China. 
The CO and its isotopologues were observed in the 1--0 transition, as well 
as several other molecular lines.
The observations were performed with a nine pixel SIS receiver,
which was configured with dual sideband (2SB) mixers for each pixel \citep{Shan10}. 
For each sideband and each
pixel, fast fourier transform (FFT) spectrometers produce a bandwidth of 1
GHz and 16384 channels, resulting in a velocity resolution of $\sim$0.16 km
s$^{-1}$ and a velocity coverage of $\sim$ 2700 \kms\ at 110 GHz.

The mapping observations were carried out in on-the-fly \citep[OTF, e.g.,][]{Mangum07} scan mode. A
position of 120$'$ to the north of the centre of W49A was taken as sky
reference. All maps are done with uniform Nyquist sampling, except for $\sim$2$\arcmin$ 
around the edges. The rms pointing uncertainty is estimated to be better than $5\arcsec$.
Typical system temperatures during the runs were $\sim$ 140 K at 110 GHz and
$\sim$ 270 K at 115 GHz.

All the PMO spectral line data were reduced with the
CLASS/GILDAS\footnote{http://www.iram.fr/IRAMFR/GILDAS} package developed by
IRAM. We classify the spectral quality of each spectrum by the baseline
flatness and system temperature levels. About 5\% of the spectra are discarded
due to poor baselines. Linear baselines are subtracted for each individual
spectrum. All spectra are then co-added and re-grided. 

We converted the antenna temperature ($T_{\rm A}^\star$) to the main beam
brightness temperature ($T_{\rm mb}$) scale using ${ T_{\rm mb} = T_{\rm
A}^{\star}(\eta_{\rm f}/\eta_{\rm mb}})$, where the 
ratio of the main beam efficiency $\eta_{\rm mb}$ to the forward hemisphere
efficiency $\eta_{\rm f}$ is 0.46.

For each molecular line, we combined all the calibrated spectra and re-gridded them
to construct datacubes with weightings
proportional to 1/$\sigma^2$, where $\sigma$ is the rms noise. This
routine convolves the gridded data with a Gaussian kernel of $\sim$1/3 the
telescope beamsize, yielding a final angular resolution slightly coarser than
the original beam size. 
The HPBW at the CO frequency is 58.0\arcsec~ (3.4 pc).

\subsection{SMA} \label{Obs:SMA}

We observed the central cluster of W49A (W49N) with the Submillimeter Array\footnote{The Submillimeter Array 
is a joint project between the Smithsonian Astrophysical Observatory 
and the Academia Sinica Institute of Astronomy and Astrophysics and is funded by the Smithsonian Institution 
and the Academia Sinica.} \citep{Ho04} in the 1.3-mm (230-GHz) band using the four 
different array configurations: subcompact, compact, extended, and very extended, 
one track for each. This multi-configuration approach images both small and large 
scales. The data set covers baseline lengths between  8 k$\lambda$ (corresponding to scales of 31\arcsec)
and  480 k$\lambda$ (0.5\arcsec). Figure \ref{fig:appa-sma} (appendix \ref{App-A}) shows the $(u,v)$ coverage of 
the combined data set.

The upgraded SMA correlator capabilities were used, covering $2\times2$ GHz in each of the two 
sidebands. The lower sideband (LSB) covered the sky frequency ranges from 218.29 to 220.27 GHz 
in spectral windows 48 to 25, and from 220.29 to 222.27 GHz in spectral windows 24 to 1. 
The upper sideband (USB) covered from 230.29 to 232.27 GHz in spectral windows 1 to 24, 
and from 232.29 to 234.27 GHz in spectral windows 25 to 48. 
The spectral resolution of the raw data set is $\approx1.1$ \kms. 

The visibilities of each track were separately calibrated using the SMA's data calibration program, 
MIR\footnote{https://www.cfa.harvard.edu/~cqi/mircook.html}. 
The phase, bandpass, and flux calibrators for each track are listed in Table \ref{tab:SMA-obs}. 
Imaging and basic analysis were done in CASA versions 3.3 and 3.4. 

The continuum emission was subtracted in the $(u,v)$ domain by fitting a baseline across the passband using line-free 
channels as input. The continuum and line data of each configuration were then combined into a single 
data set and imaged. The concatenated continuum data was self-calibrated, then imaged with different weights, checking that the flux 
recovered at different resolutions was the same within a few percent. 
The continuum map presented in this paper is the result of the combination of the SMA mosaic with the CSO BOLOCAM 
GPS archival map \citep{Ginsburg13}. 
The combination procedure is based on converting the single-dish map to the Fourier space, 
and then jointly inverting and cleaning the combined single-dish and interferometer visibilities. 
The $^{13}$CO and C$^{18}$O SMA mosaics were also combined with single dish maps to fill the short spacings. We used the IRAM 30m 
maps presented in \citep{Peng10,Peng13a}. 
The combining procedure is described in more detail by \cite{Liu13a}. 

The 11-pointing SMA mosaic is Nyquist sampled and covers all of the HC HII regions 
detected in the cm by \cite{DePree97,DePree04}. 
All the presented SMA mosaics have been corrected for primary-beam attenuation. 
Figure \ref{fig:appa-sma} (appendix \ref{App-A}) 
shows the primary-beam response of the mosaic used to correct for attenuation in the final images.  

Archival maps from JCMT-SCUBA\footnote{The James Clerk Maxwell Telescope is operated by the Joint Astronomy Centre on 
behalf of the Science and Technology Facilities Council of the United Kingdom, the Netherlands Organisation for Scientific Research, 
and the National Research Council of Canada.} at 678 GHz (project code M97Bu89) and the Very Large Array\footnote{ 
The National Radio Astronomy Observatory is a facility of the National Science Foundation operated under cooperative agreement by 
Associated Universities, Inc.} at 8.5 GHz (project code AD324) are also used in this paper. 
We refer the reader to \cite{DiFrancesco08} and  \cite{DePree97} for the respective observational details. 

\smallskip

Further analysis of the above mentioned data sets was performed in CASA\footnote{http://casa.nrao.edu/docs/UserMan/UserMan.html}, 
MIRIAD \citep{Sault95}, GILDAS\footnote{http://www.iram.fr/IRAMFR/GILDAS/}, 
DS9\footnote{https://hea-www.harvard.edu/RD/ds9/site/Documentation.html}, Karma \citep{Gooch96}, IDL, and Python. 

\section{Presentation of the data} \label{sec:data}

We present the data zooming in from large to small scales. 

\subsection{The W49 Giant Molecular Cloud} \label{sec:PMO}

\subsubsection{CO 1--0 and isotopologues}

\begin{figure}[t]
\begin{center}
\includegraphics[angle=0,scale=0.45]{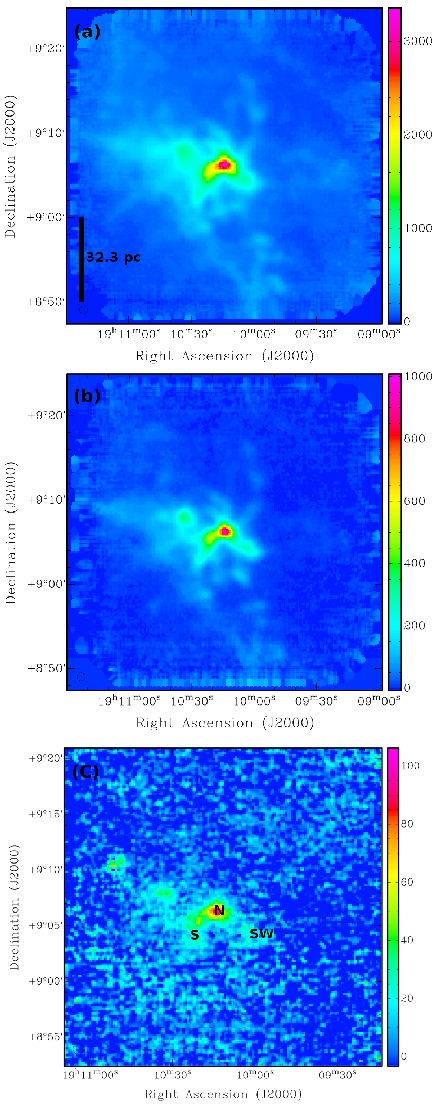}
\caption{PMO velocity-integrated (moment 0, from $-20$ to 30 \kms) CO intensity maps of the entire W49 GMC. 
Units are K \kms~. The HPBW is 58\arcsec. The scale bar of length 10\arcmin~ is equivalent to 32.3 pc. 
(\textit{a}) CO 1--0. The rms in the cleaner zones 
is $\sim 15$ K \kms,  
and is dominated by faint, extended emission. (\textit{b}) $^{13}$CO 1--0. rms $\sim 6$ K \kms.  
 (\textit{c}) C$^{18}$O 1--0. rms $\sim 6$ K \kms. The main subcomponents of W49A: W49N, W49S, and W49SW are marked in the bottom panel. 
\label{fig:W49-moms0-CO-PMO}
}
\end{center}
\end{figure}

The largest-scale maps are those obtained with the PMO 14m telescope in CO $J=1-0$ and its isotopologues $^{13}$CO and 
C$^{18}$O. These maps cover the entire W49 GMC with up to 37\arcmin~ (119 pc) per dimension. 
Figure \ref{fig:W49-moms0-CO-PMO} shows the 
velocity-integrated (moment 0) CO emission maps. This emission covers the LSR velocity range from 
$-20$ \kms~ to $30$ \kms. 
Outside this range (up to $V_\mathrm{LSR}=78$ \kms) there are more CO spikes.
Their line profiles are narrow and their emission cover the entire mapped area without matching the 
features from the W49 GMC.
We infer that they are clouds in the line of sight not associated with W49. 

The CO maps show the well-known central gas clump in the inner 3\arcmin,~ 
known as W49 North (W49N). It is in this region where the 
central embedded cluster with dozens of massive stars as traced 
by radio-continuum hypercompact (HC) and ultracompact (UC) HII regions reside \citep{DePree97,DePree04}. 
The central W49N clump appears connected to the W49 GMC via a series of filamentary extensions 
(see Fig. \ref{fig:W49-moms0-CO-PMO}). 
These filaments are hinted in previous observations where they appear as protuberances out of W49N toward the 
southwest, the east-southeast, and the north (see, e.g., fig. 2 of Nagy et al. 2012, or fig. 6 of Matthews et al. 2009). 
The southeast filament 
connects W49N to W49 South (W49S) located 2\arcmin~ apart, and which hosts a cometary HII region \citep{DickelGoss90}.

\begin{figure}[t]
\begin{center}
\includegraphics[angle=0,scale=0.7]{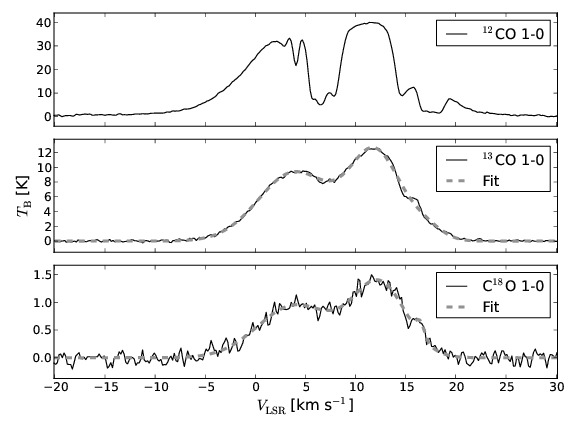}
\caption{
PMO CO spectra and Gaussian fits at the position of the emission peak, 
$\alpha\mathrm{(J2000)}=19^{\mathrm{h}}~10^{\mathrm{m}}~14.2^{\mathrm{s}},
~\delta\mathrm{(J2000)}=9^\circ~6\arcmin~23\arcsec$. The $^{12}$CO 1--0 spectrum 
({\it top}) was not fit because it is very complex, whereas 
the $^{13}$CO 1--0 and  C$^{18}$O 1--0 spectra ({\it middle} and {\it bottom}, respectively) 
are  well fitted with a sum of three Gaussians. 
\label{fig:W49-spec-CO-PMO}
}
\end{center}
\end{figure}

The spectrum from the W49 GMC has towards its position centroid two prominent velocity features. 
Figure \ref{fig:W49-spec-CO-PMO} shows the CO spectra 
toward the moment 0 peak. The optically thick $^{12}$CO is too complex to be fitted with a small number 
of Gaussians. 
The isotopologues are well fit by a sum of three Gaussian components.  
The kinematics of the full GMC and its relation to smaller scales 
will be the topic of a following paper.

\subsubsection{Other molecules and hydrogen}

\begin{figure}[t]
\begin{center}
\includegraphics[angle=0,scale=0.45]{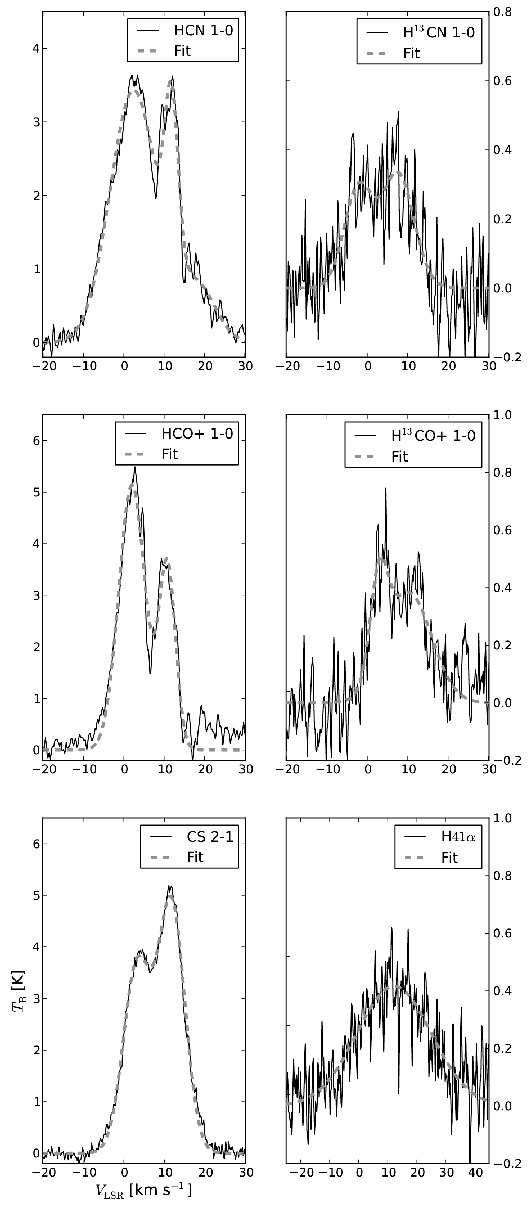}
\caption{
PMO spectra of carbon molecules (other than CO and its isotopologues) and of the H41$\alpha$ RL. 
at the position of the emission peak, 
$\alpha\mathrm{(J2000)}=19^{\mathrm{h}}~10^{\mathrm{m}}~14.2^{\mathrm{s}},
~\delta\mathrm{(J2000)}=9^\circ~6\arcmin~23\arcsec$. Gaussian fits are plotted as gray dashed lines. 
\label{fig:W49-spec-carbon-PMO}
}
\end{center}
\end{figure}

\begin{figure}[t]
\begin{center}
\includegraphics[angle=0,scale=0.45]{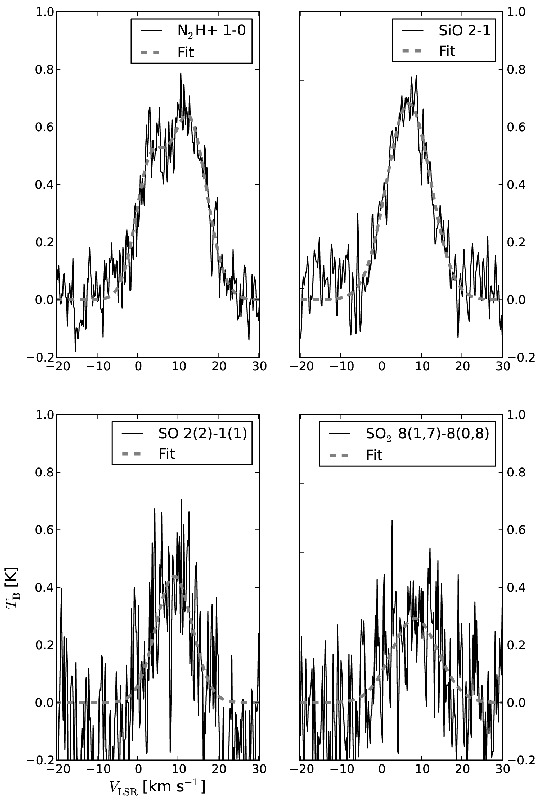}
\caption{
PMO spectra of non-carbon molecules 
and Gaussian fits at the position of the emission peak, 
$\alpha\mathrm{(J2000)}=19^{\mathrm{h}}~10^{\mathrm{m}}~14.2^{\mathrm{s}},
~\delta\mathrm{(J2000)}=9^\circ~6\arcmin~23\arcsec$.
\label{fig:W49-spec-noncarbon-PMO}
}
\end{center}
\end{figure}

Figures \ref{fig:W49-moms0-carbon-PMO} and \ref{fig:W49-moms0-noncarbon-PMO} in Appendix \ref{App-B}  
show the moment 0 (integrated from $-12$ \kms~ to $24$ \kms) of the rest of the molecules 
clearly detected in the PMO 14m observations, as well as of the hydrogen recombination line (RL) 41$\alpha$. 
Figures \ref{fig:W49-spec-carbon-PMO} and \ref{fig:W49-spec-noncarbon-PMO} show 
their respective spectra at the integrated-emission peak. 
All the transitions peak at the same position within a few arcseconds, a fraction of the beam. 
Table \ref{tab:W49-lines-PMO} lists all the lines detected with PMO and their velocity-integrated 
fluxes spatially integrated over the projected area of the GMC. The intensity conversion factor at the rest frequency 
of the CO 1--0 is 0.027 K per Jy beam$^{-1}$.

Table \ref{tab:W49-specfit-PMO} gives the parameters obtained from Gaussian fitting
to the cloud components for the different line tracers.
The bright (peak amplitude $A>1$ K) lines of HCN 1--0 and CS 2--1  
follow the CO isotopologues in having at least two clear velocity features peaked at $\sim 3$ \kms~ and $\sim 12$ \kms~
(the first peak of the CO-isotopologues is at $\sim 4$ \kms). The bright HCO+ 1--0
appears to have its two components slightly blueshifted to 2 and 11 \kms~ respectively. 
The faint isotopologues H$^{13}$CN 1--0 and H$^{13}$CO+ 
1--0, as well as the N$_2$H+ 1--0, also have two components but their parameters have larger 
uncertainties. The SiO 2--1, SO 2(2)--1(1), and SO$_2$ 8(1,7)--8(0,8) were best fit 
by a single Gaussian. 
The signal-to-noise in the latter two is too low to distinguish two Gaussians. 
The ionized hydrogen as traced by the H$41\alpha$ RL appears single peaked at $12$ \kms. 
As expected, this line is 
far broader than the molecular lines (FWHM$\approx 30.2$ \kms). 

Although we use 
several Gaussians to quantify the line features, we do not interpret them as evidence 
for separate clouds. From the CO spectra it is clear that most of the double-peaked feature is caused 
by self absorption due to the very high column densities in the line of sight to the cloud center and 
around $\sim 7$ km s$^{-1}$. This is confirmed by our calculations to create the surface density maps of 
Section \ref{surface-maps}. However, there is also a velocity gradient across W49N that contributes to create 
the line asymmetry.






\subsection{The Scale of the Central Clusters: W49N and W49S}

At the center of the W49 GMC lies the well known massive star formation region W49A, the most luminous in the 
Milky Way. W49A has two main clusters separated by $\sim 2.5\arcmin$ ($\sim 8$ pc): W49N \citep[the most 
prominent with dozens of deeply embedded (possibly still forming) O-stars as traced by UC and HC HII regions][]{DePree97}, and W49S (to the southeast). 
These scales are mapped at subarcsecond angular resolution, but with 
sensitivity to scales as large as $\gtrsim 20 \arcsec$ with the all-configuration 
SMA mosaic covering the inner $\sim2.5\arcmin$. 
The continuum mosaic, as well as the $^{13}$CO and C$^{18}$O 2--1 mosaics are combined 
with single-dish maps to recover the missing extended emission. Archival single-dish (sub)mm maps, as well 
as Very Large Array (VLA) cm maps that cover both W49N and W49S are also used. 

\subsubsection{The Continuum Emission} \label{continuum}

\begin{figure}[t]
\begin{center}
\includegraphics[angle=0,scale=0.8]{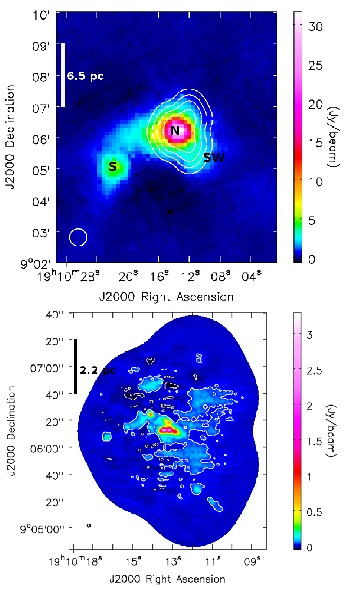}
\caption{{\it Top:} BOLOCAM GPS 1.1-mm continuum map ({\it color}) overlaid on {\it contours} 
at $\times0.2,$ 0.4, 0.6, and 0.8 the peak primary-beam response of the SMA mosaic on 
the main cluster (W49N). The second brightest peak 2.5' southeast of W49N is the secondary 
cluster usually referred to as W49 South (W49S). HPBW$=33\arcsec\times33\arcsec$. The rms noise is $\sim 40$ mJy. 
The peak intensity is 31.7 Jy beam$^{-1}$. 
{\it Bottom:} SMA mosaic combined with the BOLOCAM single-dish map. 
The HPBW of the jointly cleaned image is $2.50\arcsec\times2.26\arcsec$, PA$=75.7^\circ$. The rms noise is $\sim 5$ mJy. 
The peak intensity is $3.30$ Jy beam$^{-1}$.
\label{fig:BolocamSMAPB}}
\end{center}
\end{figure}

Figure \ref{fig:BolocamSMAPB} {\it top} shows the archival CSO BOLOCAM Galactic Plane Survey image 
\citep{Aguirre11,Ginsburg13} at 1.1 mm of 
the central $\sim25\times25$ pc of the W49 GMC. It covers W49N, W49S  
and filamentary extensions in between and toward them. 
The total flux in the BOLOCAM image is 101.0 Jy. 

It is known that HC and UC HII regions can have rising spectral indices 
($\alpha_\mathrm{ff}$, where the flux $S_\nu \propto \nu^\alpha_\mathrm{ff}$) 
in their free-free emission from cm- all the way up to mm wavelengths, with values $\alpha_\mathrm{ff}\sim 0.5$. This is due to their 
high densities and non-zero optical depths \citep{KZK08, GM09,Peters10b}.  
Therefore, to obtain dust masses when young HII regions are present it is 
necessary to subtract the free-free contribution to the mm flux. 
We use the 3.5-cm (8.5 GHz) and 7-mm (45.5 GHz) VLA data presented by \cite{DePree97,DePree04} to separate the 
total ionized-gas emission from the total dust emission. The 3.5-cm map covers both W49N and W49S and matches 
the BOLOCAM map. The 7-mm map covers only W49N where most of the HC sources reside. 
The total free-free flux at 8.5 GHz is 29.7 Jy. The compact sources have an effective $\alpha_\mathrm{ff}=0.2$, and 
contribute with 4.4 Jy of flux at 7 mm and 3.2 Jy at 3.5 cm, only 11\% of the total flux at the larger wavelength. 
We take an effective $\alpha_\mathrm{ff}=0$ for the free-free flux from 3.5 cm to 1 mm. Therefore the contributions at 1 mm 
are $\sim 29\%$ free-free, and $\sim 71\%$ dust (71.3 Jy in the BOLOCAM map). 
Feasible values of the effective $\alpha_\mathrm{ff}$ of the compact+extended emission are limited between 
-0.1 (optically thin emission) and 0.1 (a few compact sources may be outside W49N), then the limits to the dust contribution at 1 mm 
are 58\% to 79\%.

We now proceed to estimate the total gas mass in the central clusters (W49N+W49S) from the dust emission. For this we use equation  
\ref{eq:dustmass} of Appendix \ref{App-D} and a gas-to-dust mass ratio of 100. Studies of the SED of W49N from the mid infrared 
to the millimeter have found dust temperatures $T_\mathrm{d}$ from 20 K to 50 K \citep{RobsonWT90,Sievers91}. 
We assume an effective $T_\mathrm{d}=30$ K. For $T_\mathrm{d}$ 
in the extremes of the above mentionted range, 
the derived masses change by about $+70\%$ and $-45\%$. 
However, the main uncertainty in deriving masses from dust emission comes from the 
assumed dust absorption coefficient $\kappa_\nu$. 
For high-mass star formation regions, values in the range from 0.1 to 0.5 cm$^2$ g$^{-1}$ 
(per unit mass of dust) at 1.1-mm  are appropriate \citep{OH94,GM10,Beuther11}. 
The total gas mass that we obtain from 1-mm dust emission 
is $M_\mathrm{gas,d}\sim 1.5\times10^5$ to $7.6\times10^5M_\odot$ for $T_\mathrm{d}=30$ K. 
The lower end of this estimation (higher dust opacity) agrees with the CO measurements presented in Section  \ref{mass-GMC}. 
The estimations are also within the range of values obtained by 
\cite{Nagy12} and \cite{Peng13a}.

\begin{figure}[t]
\begin{center}
\includegraphics[angle=0,scale=1.0]{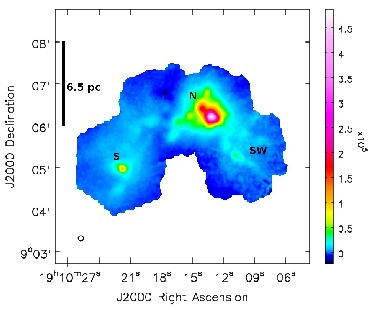}
\caption{JCMT-SCUBA map of the central part of the W49 GMC at 0.4 mm (678 GHz). The most prominent emission comes from 
the W49N cluster. W49S is also seen 2.5' southeast of W49N, as well as the filamentary structures connecting them and to the 
rest of the GMC. HPBW$=8.1\arcsec$. The rms noise is $\sim 3$ Jy. The peak intensity is 487 Jy beam$^{-1}$.  
\label{fig:SCUBA}}
\end{center}
\end{figure}

Figure \ref{fig:SCUBA} shows an 
archival JCMT-SCUBA map at 678 GHz (0.4 mm) of the central clusters W49N+W49S. 
The angular resolution of this map is 8.1\arcsec, intermediate between the combined SMA+BOLOCAM and the BOLOCAM images. 
Because dust is so much brighter at higher frequencies, this map is sensitive enough to image the filamentary structures joining 
W49N to W49S, and extending to the southwest (sometimes called W49SW), southeast, and north. 
The magnetic field in W49 has been detected by \cite{CC07} using polarized submillimeter dust continuum emission. 
The plane-of-the sky magnetic field is relatively weak (less than 0.1 mG) and well aligned along the filament joining 
W49N to W49S. 
In the rest of the paper we show that 
the filamentary structures are also seen in CO integrated intensity and column density maps, and that continue all the way up to the 
scale of the full GMC ($> 100$ pc). 

The dust mass can be independently derived from the 0.4 mm emission. 
Note that at 678 GHz and for $T_\mathrm{d}=30$ K we are completely out of the Rayleigh-Jeans regime ($h\nu/kT\sim1$), so we 
have to use the full Planck function as in equation \ref{eq:dustmass}. 
At this wavelength there is no significant contribution from free-free, and the values of the dust absorption coefficient $\kappa_\nu$ 
at shorter wavelenghts are less sensitive to the properties of dust \citep[e.g., see Fig. 5 in][]{OH94}. From \cite{OH94}, we 
consider the range of $\kappa_\mathrm{0.4mm}$ values for coagulated dust grains without thick ice mantles, 6.5 to 9.8 cm$^{2}$ g$^{-1}$. 
Then the flux in the 0.4-mm map $S_\mathrm{0.4mm}=1.96\times10^4$ Jy corresponds to a gas mass $M=4.2\times10^5$  to 
$6.3\times10^5$ M$_\odot$.

We now describe the features of the high angular resolution SMA+BOLOCAM mosaic of W49N (figure \ref{fig:BolocamSMAPB} {\it bottom}, 
covering the region marked by contours in Figure \ref{fig:BolocamSMAPB} {\it top}). 
The combining procedure is based on creating a visibility data set from the single-dish map that then is combined with 
the interferometer visibilities, fourier-transformed back to the image plane, and jointly cleaned, as described 
in \cite{Liu12b,Liu13a}. 
The combined SMA+BOLOCAM map has an angular resolution of $\sim 2.3\arcsec$. 
The subarcsecond resolution dust 
maps from the SMA mosaic only, which do not recover the extended emission but are sharper 
will be discussed along with the subarcsecond resolution SMA line mosaics in a future article. 
In this paper we focus on the mass structure 
of the W49N cluster as a whole. 

To quantify the structure in the SMA+BOLOCAM continuum map, we prefer to measure the total mass in the maps above a series  
of intensity (mass surface density) thresholds, rather than trying to extract sources using source-finding algorithms. 
The reason is that these algorithms tend to artificially fragment the emission \citep{Pineda09}. If one acknowledges the fact 
that  
the ISM is hierarchical at all scales (except possibly at $<0.1$ pc where structures will end up forming 
a single star/binary system), it follows that trying to count sources is not very meaningful.
We created masked maps with intensities $>5$ mJy beam$^{-1}$ $\times5,10,20,40,80,160,$ and 320 
(the peak in the map is 3.302 Jy 
beam$^{-1}$) and measured their total flux. 
The positive contour in Figure \ref{fig:BolocamSMAPB} {\it bottom} is at 25 mJy beam$^{-1}$. 
Table \ref{tab:flux-sma} lists the results. The quoted values for surface densities and dust-derived
masses correspond 
to $\kappa_\mathrm{1mm}=0.25$ cm$^2$ g$^{-1}$, and the errors to the results from varying $\kappa_\mathrm{1mm}$ 
from 0.1 to 0.5 cm$^2$ g$^{-1}$. As mentioned before, variations in $T_\mathrm{d}$ introduce an extra 
uncertainty (see Appendix \ref{App-D} for details). We have subtracted a free-free contribution of  $58~\%$\footnote{As 
derived before, 
the total free-free contribution in the entire VLA map is $\sim 21\%$. Within the area 
of the SMA+BOLOCAM map (W49N), an estimation of the free-free contribution under the same assumptions 
gives $\sim 58\%$.}.

\subsubsection{The Line Emission} \label{SMA-lines}

\begin{figure}[t]
\centering
\begin{tabular}{cc}
\includegraphics[angle=0,width=0.5\columnwidth]{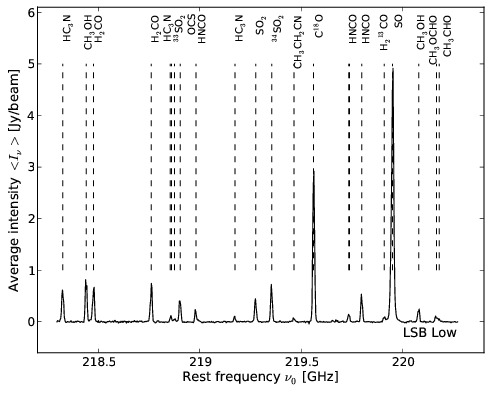} &
\includegraphics[angle=0,width=0.5\columnwidth]{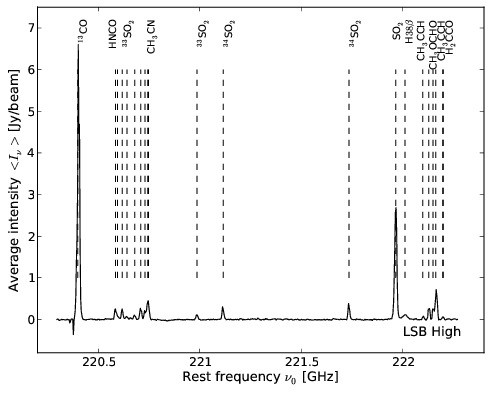} \\
\includegraphics[angle=0,width=0.5\columnwidth]{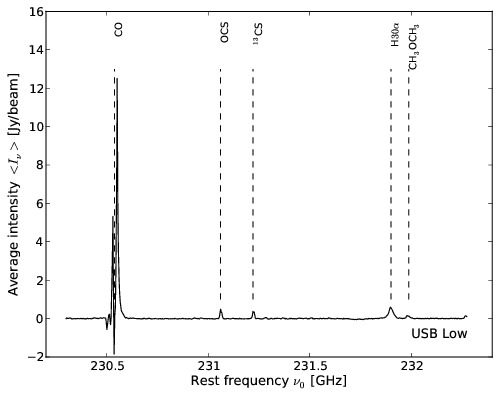} &
\includegraphics[angle=0,width=0.5\columnwidth]{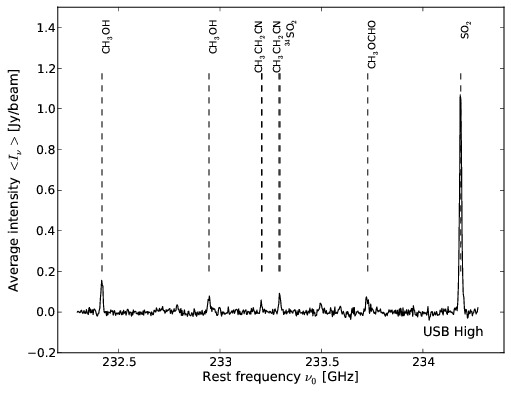} \\
\end{tabular}
\caption{
Full-sideband SMA spectra toward W49A, averaged over an area interior to the $5\sigma$ contour of the 1.3-mm 
map shown in Figure \ref{fig:BolocamSMAPB}, bottom. All the spectral features above an intensity of 50 mJy beam$^{-1}$ 
have been identified as a single or combination of spectral lines, and are marked by vertical dashed lines. 
The respective molecule or atom is labeled, except in cases of blended lines. The complete information is 
on Table \ref{tab:sma-lines}. 
\label{fig:SMAspec-fullSB}
}
\end{figure}

To identify the spectral features   
we first imaged the full bandwidth covered by the SMA using the subcompact array data 
only. Figure \ref{fig:SMAspec-fullSB} shows these spectra spatially-averaged in a contour 
matching the $5\sigma$ 1.3-mm continuum shown in Figure \ref{fig:BolocamSMAPB}, bottom. The molecules/atoms producing the
line emission are labeled. The details on the species, transition, rest frequency, and upper-level 
energy are listed in Table \ref{tab:sma-lines}. All the 57 lines that were clearly identified 
(including some blended features) above an intensity threshold of 50 mJy beam$^{-1}$ are listed. 
Many of the detected lines are typical 'hot-core' tracers  characteristic of luminous 
star formation regions \citep[e.g.,][]{Kurtz00,Cesaroni10,Beltran11}, 
although W49A appears to be less line-rich than its cousin 
Sgr B2 near the Galactic Center \citep[e.g.,][]{Nummelin98,Belloche13}. 
We have unambigously identified 
40 individual lines, plus 14 lines that are clearly identified but their profiles are blended (e.g., the CH$_3$CN 
$J=12-11$ $K=0$ and $K=1$ lines whose centers are only separated by only 5.7 km s$^{-1}$). Three more identified lines 
possibly have significant contamination (see Table \ref{tab:sma-lines}). 

W49A appears to be rich 
in SO$_2$ and its isotopologues. Younger massive star formation regions devoid of ionization or 
with only faint HII regions tend to lack emission in these lines \citep[e.g.,][]{GM10,Liu12a}, whereas regions 
in which O-stars have already formed and with HC and UC HII regions tend to be brighter in the 
SO$_2$ lines in this band \citep[e.g.,][]{GM09,Liu10}.

\begin{figure}[t]
\centering
\begin{tabular}{cc}
\includegraphics[angle=0,width=0.45\columnwidth]{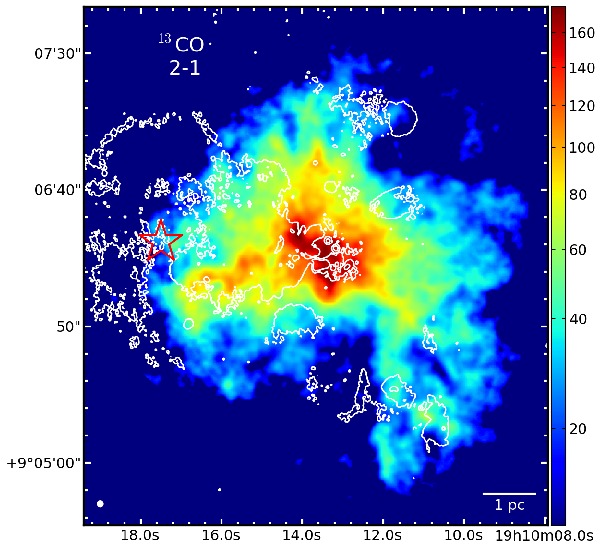} &
\includegraphics[angle=0,width=0.45\columnwidth]{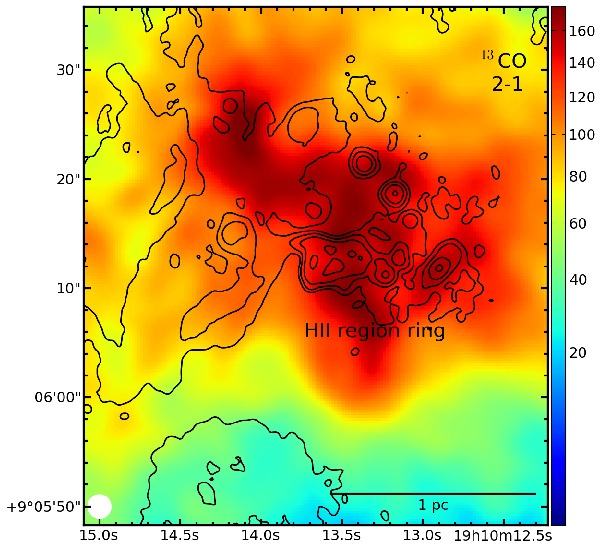} \\
\includegraphics[angle=0,width=0.45\columnwidth]{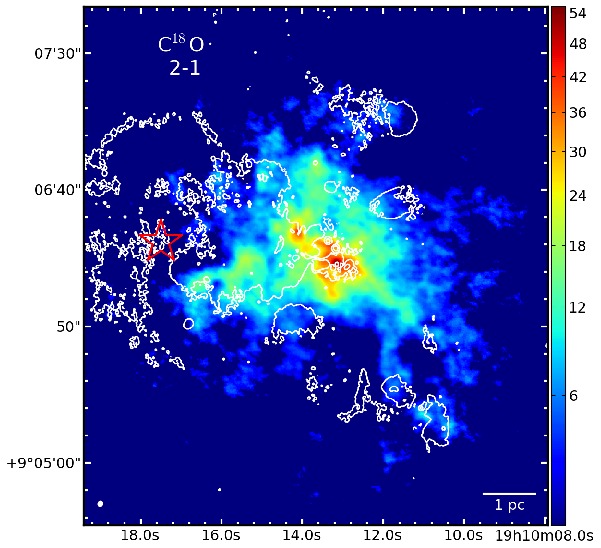} &
\includegraphics[angle=0,width=0.45\columnwidth]{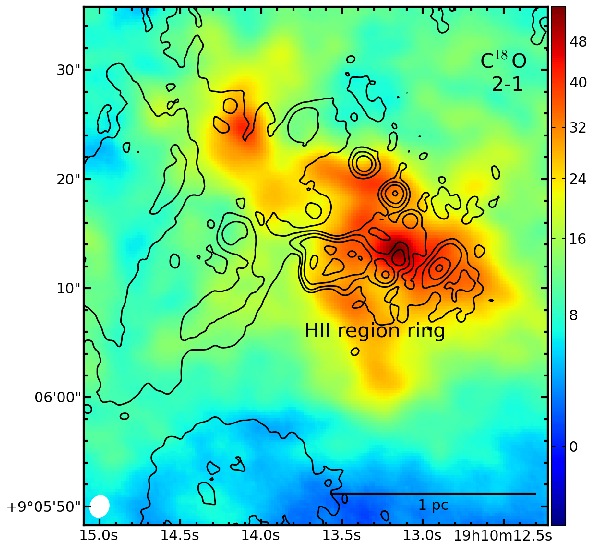} \\
\end{tabular}
\caption{Velocity-integrated (moment 0) maps of CO 2--1 isotopologues in W49A (central $\sim 10$ pc). These images show   
 the SMA (subcompact, compact, extended and very extended configurations) maps combined with IRAM 30m observations. 
The {\it color} scale shows the CO emission in Jy beam$^{-1}$ km s$^{-1}$. {\it Contours} show the 8.5 GHz (3.6 cm) continuum 
tracing free-free emission from the embedded HII regions. VLA HPBW=$0.80\arcsec\times0.78\arcsec$, PA=$-63.1^\circ$.  
{\it Top left:} Zoomed-out $^{13}$CO 2--1 covering the entire SMA mosaic. 
HPBW=$2.17\arcsec\times2.13\arcsec$, PA=$-72.1^\circ$. rms$\sim  4.8$ Jy beam$^{-1}$ km s$^{-1}$. 
The massive cluster detected in the NIR by \cite{HomeierAlves05} is marked by the red star. 
{\it Top right:} Zoomed-in  $^{13}$CO 2--1 around the central "Welch" ring of UC and HC HII regions. 
{\it Bottom left:} Zoomed-out C$^{18}$O 2--1. HPBW=$2.01\arcsec\times1.71\arcsec$, PA=$68.3^\circ$. 
rms$\sim 0.9$ Jy beam$^{-1}$ km s$^{-1}$. 
{\it Bottom right:} Zoomed-in C$^{18}$O 2--1. 
\label{fig:CO-IRAMSMA}
}
\end{figure}

After line identification, individual line cubes with the all-configuration SMA data were created. 
The synthesized beam of these maps varies smoothly from $1.02\arcsec\times0.69\arcsec$, PA$=76.8^\circ$ at the lowest 
frequency (36.4 K per Jy beam$^{-1}$) to $0.98\arcsec\times0.64\arcsec$, PA$=74.1^\circ$ at the highest 
(31.5 K per Jy beam$^{-1}$). The resolution of the final maps was chosen such that nearly all of the flux of  
the subcompact-array maps is recovered but at the highest-possible angular resolution. The CO, however, is 
affected severely by missing flux in interferometric observations. We have therefore combined the multiconfiguration 
SMA maps with IRAM 30m maps for the $^{13}$CO 2--1 and C$^{18}$O 2--1.  
Figure \ref{fig:CO-IRAMSMA} shows the velocity-integrated (moment 0) of these two CO isotopologues. As with the larger-scale PMO data, 
the emission has been integrated in the LSR velocity range from  $-20$ \kms~ to $30$ \kms. The angular resolution of 
the single-dish+interferometer maps is $\approx2\arcsec$. 

A striking feature of the SMA+IRAM30m maps is that the radially-converging, triple filametary structure seen at the full GMC scales 
(from $\sim 10$ to 100 pc) as traced by the PMO maps (section \ref{sec:PMO}) is preserved all the way down to the inner few pc. In Section 
\ref{mass-GMC} we discuss 
in more detail the morphological matching between these structures and its physical interpretation. 
Their dynamics will be discussed on a following paper. 

Appendix \ref{App-C} (on-line only) shows the velocity-integrated (moment 0) line emission of all the SMA-detected lines listed in Table 
\ref{tab:sma-lines}. A single map is shown for neighbouring lines whose profiles are blended. The last 
column of Table \ref{tab:sma-lines} lists the velocity-integrated line fluxes. 
For the science discussion of this paper (sections \ref{mass-GMC} and \ref{sec:disruption}) we use the combined SMA+IRAM 30m CO maps
together  with the dust maps and the larger-scale PMO CO maps to make 
a study of the mass of the W49 GMC as a function of scale. 

\section{The mass structure of the W49 GMC} \label{mass-GMC}

\subsection{Mass surface density maps} \label{surface-maps}

We calculate maps of the mass of the GMC from scales of $\sim 1$ pc to $>100$ pc using the PMO CO maps on the larger 
scales and the SMA+IRAM 30m CO maps on the smaller scales. The methodology is described in Appendix \ref{App-D}, and 
it is based on solving for the optical depth in every voxel of data from the CO-isotopologue ratios. 
Uncertainty in this mass determination comes from uncertainties in the isotopologue abundances (typically $\pm 50\%$), 
in the CO to H$_2$ conversion factor, and in the assumed excitation temperature $T_\mathrm{ex}$ 
(typically $\pm 25\%$). All added up, masses are uncertain within a factor of 2 each way (see Appendix \ref{App-D}).
Figure \ref{fig:Nmaps} shows the resulting maps of molecular-gas surface density $\Sigma$ 
in units of M$_\odot$ pc$^{-2}$. We derive a total mass for the GMC $M_\mathrm{tot}=1.14\times10^6$ M$_\odot$. 
In the area covered by the 0.4-mm dust map (Section \ref{continuum}), the CO mass map gives a mass 
$M_\mathrm{W49N+S+SW}=4.1\times10^5$ M$_\odot$, in agreement with the lower values in the range of masses as derived from 
0.4-mm dust emission 
($M=4.2\times10^5$  to  $6.3\times10^5$ M$_\odot$).

On the full-GMC map ({\it top} frame) 
the central cluster W49N is the most prominent feature in the center. 
Radially out of it, the known triple filamentary structure of scale $\sim 30$ pc is seen connecting W49N to W49S 
(to the southeast), to W49SW (to the southwest), and extending north of W49N. 
W49S appears to be connected to a clump $\sim 4\arcmin$ ($\sim13$ pc) to the 
northeast\footnote{This clump is also seen in archival single-dish dust images and in the individual CO maps 
(figure \ref{fig:W49-moms0-CO-PMO}).}. 
W49SW appears to end in a high density peak. Another high column density 
clump is seen $\sim 5.5\arcmin$ ($\sim 18$ pc) southwest of the main W49N peak, and is joined by a north-south 
extension to the brighter filament joining W49N to W49SW. A chain of filaments extend north of W49N and all of 
them appear curved with a similar orientation. The full GMC mass distribution is clumpy, which indicates 
fragmentation also along the filaments in the periphery. All the bright UC and HC HII 
regions reside in the central main clusters, and W49N is by far the richest in ionized-gas structures 
\citep{DePree97,DePree04}. A possible future line of research would be to study star formation in detail in the rest 
of the high column-density peaks as it has been done with W49N, looking for fainter HII regions and molecular-gas cores. 

The {\it bottom} panel of Figure \ref{fig:Nmaps} shows the high-angular resolution column density map obtained from the combination of the 
CO SMA mosaics with the IRAM 30m observations presented in \cite{Peng13a}. These map cover the area marked by a 
black square on the full-GMC map. The global column density peak is right in 
the middle of the ring of HC HIIs (see Figure \ref{fig:CO-IRAMSMA} for an overplot), at the position of the HCHII 
labeled as B by \cite{DePree97}. The highest column density features ($\Sigma>10^4$ M$_\odot$ pc$^{-2}$) are indeed 
the highest volume density regions: in Figure C1 of Appendix \ref{App-C} it is seen that many 
of the molecular lines that trace the densest 
 gas ($n>10^{5}$ cm$^{-3}$) peak at the same position as the high-resolution CO column density maps. 
 
It is striking to notice that the larger scale (10 to 100 pc) structure of the cloud is preserved all the way 
to the inner $<10$ pc. This triple structure consists of one structure of 
filaments oriented toward the east-southeast (which join W49N to W49S), one structure of filaments toward the 
southwest, and filaments extending to the north. Each of the filamentary structures is seen to be composed 
of a tree of filaments in itself. 

\subsection{A Hub-filament network}

The maps reveal that the mass structure of the W49 GMC is organized in a hierarchical network 
of filaments that appear to converge in the central, densest region harboring the W49N and W49S clusters. This network 
extends all the way from the inner pc where the embedded O-type stars in W49N reside 
\citep[as traced by the dozens of UC and HC HII regions, see Figure \ref{fig:CO-IRAMSMA}, or Figure 1 in][]{DePree97}
to the $>100$ pc scales of the full GMC (see Figure \ref{fig:W49-moms0-CO-PMO}). 
It is also preserved at intermediate scales ($\sim 10$ pc), where the filaments connect W49N to W49S, and both of them 
to the surrounding GMC (figures \ref{fig:BolocamSMAPB} and \ref{fig:SCUBA}).  

This type of hierarchical, centrally-condensed network of filaments appears to be common in clouds harboring the most luminous 
($L>10^5$ L$_\odot$) star formation regions in the Galaxy. For example, the global structure of the W49 GMC appears to be 
a scaled up version of the GMC in the G10.6--0.4 massive star formation region 
\citep[$L\sim9.2\times10^5$ L$_\odot$, see Figure 3 of][]{Liu12b}. In this scenario, the densest, central regions of the GMC 
collapse first because they have shorter free-fall time ($t_\mathrm{ff} \propto \rho^{-1/2}$). Furthermore, departures 
from spherical to sheet or filamentary geometries increase the free-fall time with the aspect 
ratio of the configuration \citep{Toala12,Pon12}. Although never completely spherical, the central W49N clump is closer to sphericity 
than the larger-scale filaments\footnote{i.e., the filament-interfilament column density contratst is larger in the cloud periphery.}. 
Therefore, it is natural to expect that the central clump (scales $<10$ pc) 
with an average number density $n\sim10^4$ to $10^5$ cm$^{-3}$ can collapse and form massive stars in a time-scale of the order of a 
few times the spherical-collapse free-fall time $t\sim10^5$ yr, whereas the more filamentary, larger-scale GMC, 
even with an initial density close to that of the central clump, will collapse in a timescale $>1$ Myr.  
Figure 1 of \cite{Liu12b} shows a schematic diagram that accounts for the aspects above discussed. 

Filament networks are often also present in low- and intermediate-mass star formation regions 
\citep[][]{Gutermuth08,Myers09,Molinari10,Pineda11,Kirk13,Takahashi13}. However, it is unclear why some of them look 
like a filament-only network and some other also have prominent hub-like condensations. This could be due to 
evolution/mass differences, or due to different observational techniques\footnote{The dust continuum, 
in contrast to line observations, picks everything on the line of sight.}. 
We hypothesize that centrally condensed networks are the main morphology 
in the  most massive clusters born out of the most massive clouds like W49 or Carina \citep{Preibisch12a,Rocca13}, but 
more systematic studies are needed. In W49, for example, many previous high-resolution studies focused 
only on the central few pc of the main clusters \citep[e.g.,][]{Wilner01,Peng10}, 
whereas studies that captured the entire GMC lacked 
the sensitivity to recover the external filaments \citep[e.g.,][]{Simon01,Matthews09}. 

Hierarchical filament networks surrounding a central cluster (or small system of clusters) are also naturally expected 
from simulations of massive clouds that form massive star clusters \citep[e.g.,][]{Bonnell01,SmithLongmore09,VS09,Dale12}. 
In these simulations, the primorial GMC has density inhomogeneities (presumably caused by the process of GMC formation itself)
that give rise to filamentary structure while the full cloud contracts. These filaments can then 
converge\footnote{Although less used in the literature, we prefer the word "convergence" to "collision", since the latter implies 
clearly distinct entities, and does not apply to, for example, a case where different filamentary arms that originated 
from a single GMC converge to a point.} into the 
hubs where cluster formation occurs \citep[e.g.,][]{GM10,DuarteCabral11,InoueFukui13},
preferably at locations deep within the global gravitational potential. When feedback 
from the formed stars is included, it is usually found that it has an important effect in regulating the star formation 
efficiency \citep[e.g.,][]{Peters10,Peters11,Krumholz11,Dale12,Colin13}.

\subsection{Other interpretations}

Several interpretations   
have been put forward to explain the observations of W49N. 
\cite{Welch87} proposed a global collapse scenario based on interferometric observations of RLs of the ring 
of HC HIIs and molecular-line absorption toward their line of sight. \cite{KLM91}, based on modeling the line 
profiles resulting from a hydrodynamical simulation, showed that the observations of 
\cite{Welch87} could be explained by a global-local collapse scenario where fragmentation from the larger-scale contracting cloud 
produces denser molecular cores that themselves collapse into individual HII regions.  
\cite{Serabyn93} proposed that the multiply-peaked single-dish line profiles are due to different molecular clumps whose 
collision triggered the vigorous star formation activity. 
The morphological evidence presented in this paper shows that the most likely situation is similar to a scaled-up version 
of the \cite{KLM91} model. The filamentary $\sim 100$ pc scale GMC has fragmented and the denser W49N clump (extending 
further toward the southeast, southwest, and north) is the converging region of 
a hierarchical (triple) network of filaments that converge toward the ring of HC HIIs. If the different filaments are 
viewed as separate entities then they could be interpreted as colliding clouds. However, the observations presented 
in this paper show that they are part of a larger, common structure. 
Kinematical evidence supporting this 
interpretation will be shown in the second paper of this project. Further evidence comes 
from looking at the available single dish line profiles: the larger the optical depth, the more prominent the $\sim 7$ 
km s$^{-1}$ dip is, indicating that it is produced by self-absorption due to the large column of gas toward the center of W49N. 
Indeed, we obtain our larger optical depths from the line ratios at these positions and velocities.  

However, the ring of HC HIIs is not the entire story. There is already a NIR massive star cluster 
concentrated within 1 pc and located $\sim 3$ pc east of the Welch ring \citep[cluster 1 of][]{AlvesHomeier03}. This cluster likely 
is the ionizing source of the extended halo of free-free emission next to the HC HIIs (see Figure \ref{fig:CO-IRAMSMA}), 
whose edge matches the shell reported by \cite{Peng10} from 4.5 and 8.0 $\mu$m observations. 
These two YMCs (which may be parts at different evolutionary stages of the same YMC), 
together with the rest of the young massive stars in W49N, W49S, and W49SE, formed from the same GMC.

\subsection{Comparison to other massive clouds in the Galaxy} \label{comparison}

We now proceed to use the column density maps described in Section \ref{surface-maps} to compute the mass of the GMC 
as a function of distance from the central cluster W49N, and to compare with other massive GMCs in the Galaxy that are 
believed to be progenitors of young massive clusters (YMCs). Figure \ref{fig:MvsR} shows the result. The match between the 
calculation using the CO 1--0 ratios on larger scales overlaps remarkably well with the result on smaller scales using CO 2--1 
ratios. The mass in a region around W49N with a diameter $D = 8.0$ pc
$M \sim 1.2 \times 10^5$ M$_\odot$, equivalent to a  a mean density under the assumption of spherical symmetry of 
$\sim 9 \times 10^3$ cm$^{-3}$. Likely, the mean density is higher 
because of the filamentary geometry and substructure. Indeed, summing the CO mass in an aperture matching 
the dust filaments seen at 0.4 mm gives $M \sim 4 \times 10^5$ M$_\odot$. 
The full GMC 
on scales of $D \sim 110$ pc (diameter) reaches a mass of $M \sim 1.1 \times 10^6$ M$_\odot$. 
In Figure \ref{fig:MvsR} we 
compare the W49 GMC with the Galactic Center cloud G0.253+0.016 \citep{Longmore12a,Kauffmann13}, which has a mass $M \sim 1.3 \times 10^5$ 
M$_\odot$ in a diameter $D = 5.6$ pc. 
Also plotted are the most massive clumps of the 
G305 GMC \citep{Hindson10} and the result of the recent study of the Carina complex by \cite{Preibisch12a}. Unlike W49 and 
G0.253+0.016, the clusters in the latter two clouds are already optically visible. G0.253 and the W49 GMC also differ 
in other properties. G0.253 is almost starless \citep{Kauffmann13,RodriguezZapata13}, appears to be externally confined 
to a total size of a few pc \citep{Longmore13}, and is in a more extreme environment in the Central Molecular Zone (CMZ) of the 
Milky Way that may difficult further star formation. On the other hand, W49 has already formed a copious amount of stars 
as seen in the infrared and radio continuum. 

It is seen that GMCs like Carina or W49 have a gas mass reservoir large enough ($\sim 10^6$ M$_\odot$) and the right size (radius $\sim10^1-10^2$ pc) 
to form stellar clusters with the typical stellar mass $M_\star \sim 10^5$ M$_\odot$ and radius $\sim 10$ pc of globular clusters \citep{PZ10}. 
The gas-mass reservoir of the central few pc is still enough to 
form a YMC with a stellar mass $M_\star > 10^4$ M$_\odot$, even if it decouples 
completely from the rest of the cloud.

We note that the gas mass is dominated by molecular gas at all the scales relevant to this discussion. From the 3.5-cm free-free map, and 
for a range of turn-off frequencies\footnote{The frequency at which an HII region transitions from being optically thin to thick.} from 
8 to 230 GHz, the total HII mass is between $M_\mathrm{HII} \sim 1700$ and 5400 $M_\odot$.

\section{GMC and cluster disruption} \label{sec:disruption}

What is the future of the W49 GMC and its star clusters? Will the GMC be disrupted 
and star formation terminated? Will the resulting star cluster remain bound? 

\subsection{GMC disruption by feedback}

The GMC can be disrupted by a series of feedback processes, from which 
the most important are protostellar jets and outflows, HII pressure, and radiation pressure. 
In W49A, the luminosity output is heavily dominated by the most massive stars.
Massive, collimated outflows, do not appear to be important. 

In principle, the over-pressurized ionized gas at temperature $T\sim10^4$ K can clear out 
the surrounding molecular gas, which is at a temperature from tens to hundreds of Kelvin \citep{Keto07}. However, W49A is 
at a stage at which in spite of its enormous luminosity, several of its HII regions are still confined 
within $\sim 1000$ AU \citep{DePree00,DePree04}. Various mechanisms may be responsible for delaying HII 
region growth, including confinement by denser molecular gas \citep{DePree95}, 
molecular infall from the surrounding clump \citep{Walmsley95}, 
and continued accretion through filamentary, partially-ionized accretion flows \citep{Peters10,GM11,Dale12}. 
W49A has already a lifetime of several $\times10^5$ yr since the onset of massive star 
formation\footnote{The timespan of any previous low-mass star formation, plus the timespan of 
the assembly of the GMC are apart.}, 
given the presence of hypercompact HII regions and  hot molecular cores \citep{Wilner01}. 
After this time, only $\sim 1~\%$ of the gas mass in the inner $\sim 6$ pc around W49N is 
ionized, the rest is molecular. The difference is even more dramatic ($\sim 0.1~\%$ of gas is ionized)
if the full GMC is considered. 
Therefore, it is evident that in this timescale ionization is not efficient to disrupt the cloud. 
Even if the few central pc of gas 
were fully ionized, the main disruptive force would be radiation pressure rather than ionized-gas pressure. 
From Figure 2 of \cite{KM09}, 
for an ionizing photon rate $Q\sim$ few $\times10^{51}$ s$^{-1}$, and an average particle density 
$n\sim$ few $\times10^3$ cm$^{-3}$, the radiation pressure would be dominant by a factor $\sim 100$.

To estimate the effect of radiation pressure, since most of the gas is molecular and not ionized, we should 
consider that 
the dominant form will be onto dust 
grains\footnote{Even photons that are first absorbed by hydrogen atoms and returned at lower energies will be 
absorbed by dust \citep[e.g.,][]{KM09}.}.
The GMC is disrupted if the outward radiation pressure from the central 
clusters overcomes the cloud's self gravity, i.e., if the luminosity from the clusters $L_\star$ is greater
than the Eddington luminosity $L_\mathrm{Edd}$. 
Under spherical symmetry (an assumption only valid to some extent in the inner $\sim 10$ pc of the central 
clusters): 

\begin{equation}
L_\mathrm{Edd}(r)=4\pi cG M(r) / \kappa_\mathrm{R}, 
\end{equation}

\noindent
where $M(r)$ is the dust mass at radius $r$ and $\kappa_\mathrm{R}$ is the Rosseland mean dust opacity. 
We take $M(r=6.5\mathrm{pc})\sim2.1\times10^3$ $M_\odot$ from our CO and dust 
measurements\footnote{Note that this is the mass in the dust component.}, 
and $\kappa_\mathrm{R}$ in the range from 0.1 to 1 cm$^2$ g$^{-1}$ \citep{Semenov03}. 
Therefore, $L_\mathrm{Edd}(r=6.5\mathrm{pc})$ is in the range from $2.7\times10^7$ to $2.7\times10^8~L_\odot$.
\cite{Sievers91} obtained the luminosity of the central clusters by fitting their spectral energy 
distribution from the mid infrared to the millimeter. We rescale their result to a distance of 11.1 kpc (those 
authors used 14 kpc), to obtain a total luminosity $L_\mathrm{bol}\approx1.7\times10^7~L_\odot$. The Eddington 
ratio $L_\mathrm{bol}/L_\mathrm{Edd}$ is then in the range 0.06 to 0.6. 
Note that the radius selection corresponds
to the distance of W49N to W49S, and therefore is roughly the minimum gas mass that encloses most of the luminosity. 
If larger radii are taken, or if the larger mass estimates from taking an aperture following the central filaments in 
the CO maps, or from the 0.4-mm observations are taken, 
then the Eddington ratio $L_\mathrm{bol}/L_\mathrm{Edd}$ would be even lower. 
Also, if the totality of the GMC is considered, it is possible that the filaments that are seen to radially converge in 
the central clusters continue to feed them during the star-formation timescale. A dynamical 
study of the filamentary network will follow this paper, however, even in the static case, we can 
conclude that the current cluster is not able to entirely clear its own cloud. 

\cite{Murray10} investigated the disruption of GMCs across a wide range of conditions, from Galactic 
starburst regions like W49 to Ultra-luminous Infrared Galaxies. They indeed conclude that radiation pressure 
is the dominant force in cloud dispersal. For their W49A case, however, they pick a cluster luminosity $L=6.4\times10^7~L_\odot$
from correcting the free-free luminosity for dust absorption. This higher cluster luminosity in their model allows it 
to marginally disperse the GMC. Although the luminosity derived from the far-IR by \cite{Sievers91} is more accurate, it is also 
true that the cluster may still gain more mass and become more luminous. 
An appealing idea is that the cluster will continue 
to gain mass until $L_\mathrm{bol}/L_\mathrm{Edd}\approx1$.
Indeed, there is evidence that star formation in the 
central clusters is continuous: part of the stellar population is already visible in the near 
infrared\footnote{Concentrated $\sim 3$ pc east of the Welch ring of HC HIIs, the "Cluster 1" of \cite{AlvesHomeier03} 
likely is the ionizing source of the extended halo of ionized emission seen in the cm \citep{DePree97}, whose edge matches the 
shell reported by \cite{Peng10}.} \citep{ContiBlum02,AlvesHomeier03}; 
part of it is barely visible in the near- and mid-IR \citep{Smith09} but their massive stars are seen 
as HC and UCHII regions in the radio \citep[e.g.,][]{DePree97,DePree00}; and finally, there are also hot molecular cores without associated 
HII region \citep{Wilner01}.

\subsection{The fate of the central star cluster}

Regarding the future of the star clusters, W49N and W49S may form more than one YMC, or coalesce into a single system. 
They could also dissolve. Observations and models of YMC dynamics suggest that if the star-formation efficiency is low enough 
($< 50$ \%) "rapid"\footnote{Rapid compared to the crossing time $t_\mathrm{cr}$. 
For a GMC of size $\sim10$ pc and velocity dispersion $\sim 5$ km s$^{-1}$, $t_\mathrm{cr}\sim2$ Myr.}  
gas dispersal leaves unstable star clusters that end up dissolving \citep[][]{Lada84,BastianGoodwin06}. 
However, if the gas dispersal timescale is comparable or larger than the crossing time of the GMC, the resulting 
star cluster has more time to adjust to the new gravitational potential and it is easier for it to 
remain bound even with star formation efficiencies as low as 10  \% \citep{Lada84,PPZ12}. 

We argue that, because the star formation efficiency $SFE=M_\star/(M_\star+M_\mathrm{gas})$ and the gas dispersal timescale are large enough,  
the most likely 
outcome is that the star clusters W49N+W49S (or at least part of their stellar content) will remain bound.

From near-infrared observations, 
\cite{HomeierAlves05} estimate a stellar mass in frames of length $\sim 5\arcmin$ ($\sim 16$ pc) 
of $M_\star\sim 4\times10^4$ to $7\times10^4~M_\odot$. The total stellar mass should be somewhat larger when the embedded 
population not visible in the IR (e.g., the HC HII regions) is taken into account.
The total gas mass in the same area from the CO and submm continuum data is 
$M_\mathrm{gas}\sim 4\times10^5$ to $ 6\times10^5~M_\odot$.
Therefore, within the W49N+W49S region, the current $SFE$ is $> 10~\%$, 
and will likely increase with time until star 
formation is terminated.

Observations of the most luminous star formation 
regions in the Galaxy show that their star clusters are still deeply embedded on large scales even after the formation 
of subsequent generations of UC HIIs, HC HIIs, and hot molecular cores, which requires timescales of at least several $\times10^5$ yr 
from the onset of significant feedback \citep{GM09,Liu11,Liu12a,Liu12b}. Older, optically visible clusters like the one in 
Carina \citep{Preibisch12a} are still surrounded by hundreds of thousands of solar masses of molecular gas. Therefore, observations show that 
gas dispersal is slow compared to the star formation timescale. 
It is harder to estimate the SFR of the whole GMC because although in this paper we have presented a good account 
of its gas content, most of the stellar content is deeply embedded. 

The central part of the W49 GMC satisfies other proposed criteria for bound massive cluster formation. 
\cite{Bressert12} proposed that massive clusters need to have a escape velocity $v_\mathrm{esc}$ greater than the sound speed of 
ionized hydrogen 
$c_\mathrm{HII}\approx10$ km s$^{-1}$. The central clusters in W49A satisfy this criterion. For a current star cluster mass 
$M_\mathrm{cl}\sim7\times10^4~M_\odot$, the minimum bound-cluster radius is $r_\mathrm{cl}=2GM_\mathrm{cl}/c_\mathrm{HII}^2\approx6$ pc. 
If the embedded population is taken into account, assuming $M_\mathrm{cl}=10^5~M_\odot$, $r_\mathrm{cl}=8.6$ pc. The 
near-IR 
clusters\footnote{``Cluster 1" is associated with W49N, "cluster 2'' with W49S, and "cluster 3'' with W49SE.} 
identified by \cite{AlvesHomeier03} have individual radii $r<1$ pc. 
If further contiguous star formation ends up making 
a single cluster covering all the central part of the cloud, from W49S to W49SE, the radius of the resultant cluster is $r<6$ pc, 
compact enough to satisfy the criterion of \cite{Bressert12} for boundness. This, however, may be a very 
restrictive condition, since it assumes that even a fully ionized cloud is contained by gravity and will eventually be used to form stars. 
Our observations show that in the W49 GMC  the ionized-gas mass fraction is very small. 

\cite{Kruijssen12} presented a model in which bound star clusters are formed out of the high density end of the ISM. This model
accounts for possible dissolution effects like gas expulsion and external tidal forces, for conditions from typical 
Milky Way to extragalactic starbursts. 
\cite{Kruijssen12} finds that for gas mass surface densities above $\Sigma_\mathrm{gas} \sim 1000$ M$_\odot$ pc$^{-2}$, 
about 70 \% of the formed clusters will remain bound. Our surface density map (figure \ref{fig:Nmaps}, bottom right) shows that 
almost all of the W49N gas is above this threshold. This further suggests that the central cluster W49N\footnote{We do not have a 
high-resolution mosaic of W49S.} will remain as a bound one.

\section{Conclusions} \label{sec:concl}

The first results from our MUSCLE survey of gas in the W49 GMC can be summarized as follows: 

\begin{itemize}
\item 
The W49 GMC is one of the most massive in the Galaxy.  
From multi-scale observations of CO and isotopologues, 
we derive a total mass $M_\mathrm{gas}\sim1.1\times 10^6$ M$_\odot$ within a radius of 60 pc.
Around the most prominent cluster W49N forming at the center of the GMC, within a  radius of 6 pc, the total gas mass is 
$M_\mathrm{gas}\sim2\times 10^5$ M$_\odot$ (masses uncertain to $\pm 50\%$). 
Therefore $\sim 20~\%$ of the gas mass is concentrated in $\sim 0.1~\%$ of the volume. 
The gas mass is dominated by the molecular- rather than the ionized phase 
(only $\sim 1~\%$ of gas is ionized in the inner region). The W49 GMC has 
enough mass to form a young massive cluster (YMC) as massive as a globular cluster with a conservative star 
formation efficiency. 
We compare our results with recent studies of clouds candidate to form YMCs, like the Galactic Center cloud G0.253+0.016 
\citep{Longmore12a}, the Carina complex \citep{Preibisch12a}, and G305 \citep{Hindson10}.

\item
The mass of the GMC is distributed in a hierarchical 
network of filaments that is forming a young massive cluster (YMC), or a system of YMCs. 
At scales $<10$ pc, a triple, centrally condensed, 
filamentary structure peaks toward the ring of hypercompact HII regions in W49N known to host dozens of deeply embedded 
(maybe still accreting) O-type stars. This structure is observed to continue at scales from $\sim 10$ to 100 pc through 
filaments that radially converge toward W49N and toward its less prominent neighbour W49S. These large scale filaments are clumpy and 
could be forming stars at a rate lower than that of the central clusters. This finding suggests that the W49A starburst most likely formed 
from global gravitational contraction with localized collapse in a "hub-filament'' geometry.  

\item
Feedback from the central YMCs (with a current mass $M_\mathrm{cl} \gtrsim 7\times10^4$ M$_\odot$) 
is still not enough to disrupt the GMC, but further stellar mass growth within a factor of 2 could be enough to allow radiation 
pressure to disrupt the cloud and halt star formation. There is no evidence on global scales for significant disruption 
from photoionization. 

\item
Likely, the resulting stellar content will remain as a gravitationally bound massive star cluster, or a small system of bound star clusters. 

\end{itemize}

\acknowledgments
The authors acknowledge an anonymous referee for a timely and useful report. 
The authors are grateful to Mr. Bing-Gang Ju and Mr. Deng-Rong Lu from the PMO operations staff. 
R.G.-M. acknowledges funding from the European
Community's Seventh Framework Programme (/FP7/2007-2013/) under grant
agreement No. 229517R. L.F.R. and L.Z. acknowledge the support of DGAPA, UNAM,
and CONACyT (Mexico). T.P. acknowledges financial support through
SNF grant 200020\textunderscore 137896. R.G.-M. and E.K. acknowledge the hospitality of the Aspen Center for Physics, 
which is supported by the National Science Foundation Grant No. PHY-1066293.
This research made use of APLpy, an open-source plotting package for Python hosted at http://aplpy.github.com. 
The authors acknowledge the anonymous referee for an encouraging and useful report. 
R. G.-M. thanks Quang Nguyen Luong and Diederik Kruijssen for comments on a draft of the paper.
J.E.P has received funding from the European Community's 
Seventh Framework Programme (/FP7/2007-2013/) under grant agreement No 229517 and from the SNF 
(Swiss National Science Foundation) Sinergia Project.

\bibliographystyle{apj}
\bibliography{biblio}

\newpage

\appendix

\section{A: Details of the SMA mosaics} \label{App-A}

The SMA observations make use of the 4 available array configurations to cover the baseline range from 8 to 480 k$\lambda$ 
(Fig. \ref{fig:appa-sma}, left). 
Interferometric observations of massive star formation regions (and in general, of fields with complicated structure at 
multiple scales) are typically limited by the dynamic range produced by the discrete $u,v$ coverage rather than by integration 
time as in point-source observations. We have corroborated in our data set, as well as in other studies over the past years 
\citep[e.g.,][]{GM10} that co-adding SMA compact and extended configurations helps to recover all the flux up to scales of $\sim 10''$
(comparable to the core scales of $\sim 0.1$ pc at typical kpc distances) while preserving subarcsecond angular resolution. 
However, since the ISM is hierarchical, cores are themselves embedded in pc scale clumps which may themselves be embedded on 
larger-scale structures, as shown in this paper. To map these structures, combined single-dish and inteferometer mapping 
is needed. ALMA is supposed to resolve these issues by combining mapping from two different arrays and single dishes. 
The data presented here covers a larger range of angular scales (plus the single dish) than 
even what has been offered for ALMA Cycle 1. 

The right panel of Figure \ref{fig:appa-sma} shows the primary beam response of the 11-pointing mosaic. The images and measurements 
made in this paper have been divided by this reponse to correct the flux scale. Most of the interferometer  
maps have most of the flux in the central part of the mosaic. However, the CO (and isotopolgues) and dust continuum maps suffer from missing 
flux even in the all-configuration mosaics, so we have combined them with single-dish data. 
The SMA continuum map has been combined with the BOLOCAM GPS survey map \citep{Ginsburg13}. 
The $^{13}$CO and C$^{18}$O 2--1 maps have been combined with IRAM 30m maps presented in \cite{Peng13a}. The combining procedure 
is described in \cite{Liu12b,Liu13a}. It is based on creating a visibility data set from the single-dish maps that overlaps in 
$u,v$ range with the interferometer data. The concatenated data is then inverted back to the image space and jointly cleaned.

\section{B: PMO velocity-integrated maps} \label{App-B}

In this Appendix we show the velocity-integrated (moment 0) PMO maps of the lines listed in Table \ref{tab:W49-lines-PMO}, except for 
the CO maps that are used in the body of the text.

\section{C: SMA velocity-integrated mosaics} \label{App-C}

In this Appendix we show the velocity-integrated SMA maps from all the identified lines listed in Table \ref{tab:sma-lines}, 
except for the $^{13}$CO and C$^{18}$O which are used in the main body of the text. 
A detailed analysis of these maps will be presented in a forthcoming paper.

\section{D: Calculations of the mass of the GMC} \label{App-D}

We calculate the mass of the W49 GMC at all scales using four data sets: the combined SMA+IRAM 30m 
$^{13}$CO and C$^{18}$O 2--1 maps of the inner $\sim 10$ pc (Fig. \ref{fig:Nmaps}), 
the combined SMA+BOLOCAM GPS 1 mm 
dust continuum map also in the inner $\sim 10$ pc (Fig. \ref{fig:BolocamSMAPB}), 
the archive SCUBA 0.4-mm map (Fig. \ref{fig:SCUBA}), 
and the CO and $^{13}$CO 1--0 PMO maps covering the full 
GMC up to $\sim 110$ pc scales (Fig. \ref{fig:Nmaps}).  Below we describe the methodology. 

\subsection{CO 1--0}

If we assume that the $^{13}$CO emission is optically thin, the $^{12}$CO opacity ($\tau_\mathrm{^{12}CO}$) in a given voxel of 
data (a position-position-velocity cell) can be calculated from the intensity (brightness) ratio of the lines
\citep{Snell84}: 

\begin{equation}
\frac{T_\mathrm{B,^{12}CO}}{T_\mathrm{B,^{13}CO}}=\frac{1-\exp(-\tau_\mathrm{^{12}CO})}{1-\exp(-\tau_\mathrm{^{12}CO}/\chi_{^{13}C})}, 
\end{equation}

\noindent 
where $\chi_{^{13}C}$ is the relative abundance of $^{12}$C to $^{13}$C. 
We use  $\chi_{^{13}C}=65$ using the fit of \cite{WilsonRood94} 
for a Galactocentric distance of 7.6 kpc for W49, derived using its parallax distance to the Sun of  11.1 kpc 
\citep{Zhang13}. 

In the voxels where there is a $^{12}$CO detection but no $^{13}$CO we use the optically thin approximation to the 
column density of $^{12}$CO [cm$^{-2}$]: 

\begin{equation}
N_\mathrm{thin,^{12}CO}=4.31\times10^{13} \frac{T_\mathrm{ex}+0.9}{\exp(-5.5/T_\mathrm{ex})} T_\mathrm{B,^{12}CO} \Delta v , 
\end{equation}

\noindent 
where temperatures are in K and velocity widths in km s$^{-1}$. 

In the voxels with $^{13}$CO emission we calculate the optically-thick column density with: 

\begin{equation}
N_\mathrm{thick,^{12}CO}=\frac{\tau_\mathrm{^{12}CO}}{1-\exp(-\tau_\mathrm{^{12}CO})} N_\mathrm{thin,^{12}CO}. 
\end{equation}

Once the column density is known for each position-position-velocity voxel, we integrate them to obtain the mass surface
density maps shown in Section \ref{mass-GMC}. 
We note that in principle the same procedure can be used with the C$^{18}$O to $^{12}$CO ratio. However, the PMO C$^{18}$O maps 
are not sensitive enough to detect emission except for the very center of the cloud (see Section \ref{sec:PMO}). 

Two small corrections have been applied to the straightforward procedure outlined above. 
First, there is a dip in the spectra between 14 and 18 km s$^{-1}$ that appears to be caused by an extended, 
foreground cloud \citep{Peng13a}. We have interpolated the measurements within this range. The corrected mass is 
$<1 \%$ larger. Second, a few voxels close to the map center and around $\sim 7$ km s$^{-1}$ have very large optical depths 
(i.e., the $^{12}$CO and $^{13}$CO maps have the same intensity within the uncertainty). 

For a cloud that is fully molecular, a hydrogen to helium number ratio of 10, and an abundance of H$_2$ to $^{12}$CO $\chi_{CO}=10^4$,  
the gas mass surface density $\Sigma$ is then obtained from: 

\begin{equation}
\Sigma=\mu m_\mathrm{H} \chi_\mathrm{CO} N_\mathrm{^{12}CO}, 
\end{equation}

where [$\Sigma$] = g cm$^{-2}$ for [$N_\mathrm{^{12}CO}$] =  cm$^{-2}$, $\mu=2.8$, and $m_\mathrm{H}=1.6733\times10^{-24}$ g.

\subsection{CO 2--1}

On the scales of W49N ($\sim 10$ pc) we follow a similar procedure to the one outlined above, but with the $^{13}$CO  and 
C$^{18}$O 2--1 maps at $\sim 2''$ angular resolution combined from the SMA mosaics and IRAM 30m data. In this case, we have that: 

\begin{equation}
N_\mathrm{thin,^{13}CO}=1.17\times10^{13} \exp(5.3/T_\mathrm{ex}) \frac{T_\mathrm{ex}+0.9}{\exp(-10.6/T_\mathrm{ex})} T_\mathrm{B,^{13}CO} \Delta v. 
\end{equation}

Th rest of the procedure is as in the larger-scale maps. 
The relative abundance of $^{13}$CO to C$^{18}$O (7.5) is the ratio of the abundance of 
$^{16}$O to $^{18}$O ($\chi_{^{18}O}=484.3$) 
to the abundance of  $^{12}$C to $^{13}$C  ($\chi_{^{13}C}=65$) \citep{WilsonRood94}. 
The maximum optical depth $\tau_\mathrm{C^{18}O}$ solved in our 
calculations is 20. As with the larger-scale GMC, we have interpolated the spectra within the velocity range 
from  14 and 18 km s$^{-1}$.

\subsection{Millimeter continuum}

Dust grains can be assumed to emit as a gray body at temperature $T_\mathrm{d}$. At millimeter and submillemeter wavelenghts their emission 
is optically thin, so the total mass in dust grains is given by 

\begin{equation}
M_\mathrm{d}=3.25\times10^6 d^2 \biggl ( \frac{\exp(0.048\nu / T_\mathrm{d}) -1}{\nu^3} \biggr ) \frac{S_\nu}{\kappa_\nu},  
\label{eq:dustmass}
\end{equation}

\noindent
where the units of mass [$M_\mathrm{d}$], distance [$d$], frequency [$\nu$], dust temperature [$T_\mathrm{d}$], flux [$S_\nu$], 
and dust absorption coefficient [$\kappa_\nu$] are respectively M$_\odot$, kpc, GHz, K, Jy, and cm$^2$ g$^{-1}$. 

The total molecular-gass mass is then $M_\mathrm{gas}=100M_\mathrm{d}$. Often the Rayleigh-Jeans regime ($h \nu \ll k T_\mathrm{d}$) is 
assumed, but this is not the case for $T_\mathrm{d}\sim30$ K and the frequencies we deal with, specially at 678 GHz (0.4 mm) where 
$h \nu / k T_\mathrm{d} \approx 1$. Therefore we prefer to use the full Planck function.

\subsection{Uncertainties} 

Many sources of uncertainty make the estimation of masses derived from CO ratios and dust continuum emission accurate 
to a factor of 2, at most. 
For the CO-ratio measurements, we handle uncertainties on the isotopologue abundances and the excitation temperature, 
and do not consider uncertainties in the CO to H$_2$ abundance factor, which is usually set in the literature 
to $\chi_{CO}=10^{4}$. 
For the dust measurements, we handle uncertainties in the dust absorption coefficient and the dust temperature, 
but again leave the gas-to-dust mass conversion factor at 100. 

The uncertainty in the mass determinations from CO isotopologue abundances is estimated from the 
fits of \cite{WilsonRood94}. The abundance of $^{12}$C with respect to $^{13}$C is $\chi_{^{13}\mathrm{C}}=65\pm27$, and the abundance 
of  $^{16}$O to $^{18}$O is $\chi_{^{18}\mathrm{O}}=484\pm172$. For the optically-thin and the non-saturated 
optically thick voxels, uncertainty in the mass propagates as the uncertainty in abundance. However, this is not the case
 for the saturated voxels in which 
we set $\tau$ to a maximum value $\tau_\mathrm{max}$. $\tau_\mathrm{max}$ is defined as the value at which 
the line brightness ratio equals $1+\sigma$, where $\sigma$ is the rms noise in the maps, i.e., larger optical depths are not 
distinguishable, within the noise. We have found empirically how this affect our calculations by re-running 
them within the allowed range of abundance values. In the large-scale PMO maps, very few pixels are affected by saturation in all cases. 
Therefore the masses scale almost linearly with the used abundance. In the smaller-scale SMA maps, several voxels are saturated 
mainly in the central parts of the map, therefore larger abundances permit increasingly large  $\tau_\mathrm{max}$ and larger mass estimates. 
The mass uncertainty due to the above mentioned effect grows at smaller radii, up to $+130\%$ and $-57\%$ within $r=0.5$ pc. 

If the excitation temperature $T_\mathrm{ex}$ is not position dependent, 
then its effect on the uncertainty of the mass estimate is almost linear and affects equally 
all the radial measurements. We pick $T_\mathrm{ex}=30$ K. Reasonable $T_\mathrm{ex}$ values are in the range from 20 to 40 K. 
Therefore we estimate a respective mass uncertainty of $\pm 26\%$. We also explore the effect of a possible radial 
dependence $T_\mathrm{ex} \propto r^{-1.5}$, with $T_\mathrm{ex}(r=0)=50$ K, and  $T_\mathrm{ex}(r=\mathrm{5 pc})=20$ K 
(i.e., a warm W49N). The results 
show that the mass in the inner 5 pc would be systematically underestimated 
by $\sim 54\%$ close to the center and  $\sim 27\%$ at $r=5$ pc, and then at larger radii the mass difference would converge to 
$\sim 3\%$ at $r=50$ pc. However, we do not include the effects of this hypothetical temperature profile in our error budget.

Propagating the uncertainties from abundances and excitation temperatures gives a total uncertainty of $\sim \pm 50\%$ for the PMO 
masses. For the SMA+IRAM 30m masses, typical values are $+80\%$ and $-50\%$, but uncertainties are larger at smaller radii. 
The error bars in Figure \ref{fig:MvsR} include the individual contributions from abundance uncertainty and from shifts 
in $T_\mathrm{ex}$.

\smallskip

Uncertainties in mass determinations from dust are as large or larger than from CO. The dust absorption 
coefficient $\kappa_\nu$, dependent on the properties of dust grains, is the main source of uncertainty. We take as 
possible values for $\kappa_\nu$ those corresponding to coagulated dust grains without significant ice mantles from 
the calculations of \cite{OH94}. The dispersion of these values is smaller at 0.4 mm than at 1 mm. We estimate the mass 
uncertainty from $\kappa_\nu$ to be $\pm \sim 20\%$ at the former wavelength, and $\pm \sim 65\%$ at the latter. 
Possible variations from the adopted dust temperature $T_\mathrm{d}=30$ K also contribute to the total uncertainty in mass. 
From the gray body fits of \cite{Sievers91} and \cite{RobsonWT90}, $T_\mathrm{d}$ is restricted to the range between 20 and 50 K. 
Therefore the uncertainty from temperature in the mass determination is $\sim \pm 50\%$. 
Again, we do not consider further uncertainties from the dust-to-gas mass conversion factor, which we fix at 100. 
Therefore, we estimate a total mass uncertainty of $\pm 55\%$ at 0.4 mm, and $\pm 90\%$ at 1 mm, including uncertainties 
from the free-free subtraction at the latter wavelength.

\clearpage

\clearpage

\clearpage

\clearpage

\clearpage

\clearpage

\clearpage

\clearpage

\clearpage

\begin{figure}[t]
\centering
\includegraphics[angle=0,scale=0.30]{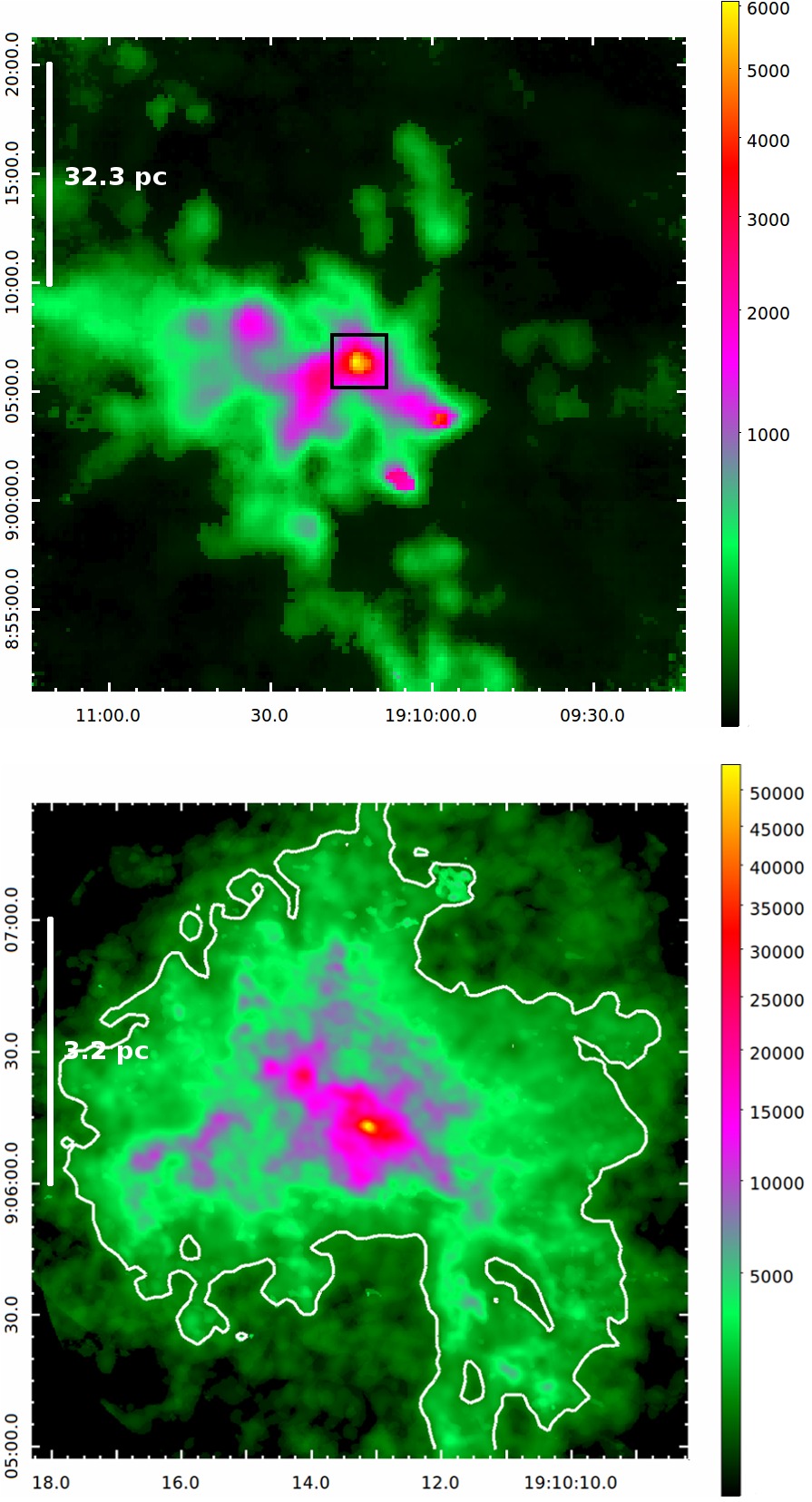}
\caption{Mass surface density $\Sigma$ maps obtained from CO-isotopologue line ratios, as described in Appendix \ref{App-C}. 
Units are M$_\odot$ pc$^{-2}$. 
The {\it top} frame shows the zoomed-out $\Sigma$ measurement from the PMO CO and $^{13}$CO 1--0 maps. HPBW=$58\arcsec$. 
The {\it bottom} frame shows the zoomed-in $\Sigma$ measurement from the SMA mosaics combined with IRAM 30 maps of $^{13}$CO and 
C$^{18}$O 2--1, covering the area marked by a black square in the top frame. The {\it contour} in the bottom frame corresponds 
to 1000 M$_\odot$ pc$^{-2}$. HPBW=$2.2\arcsec\times2.2\arcsec$.
}
\label{fig:Nmaps}
\end{figure}

\clearpage

\begin{figure}[t]
\centering
\includegraphics[angle=0,scale=0.9]{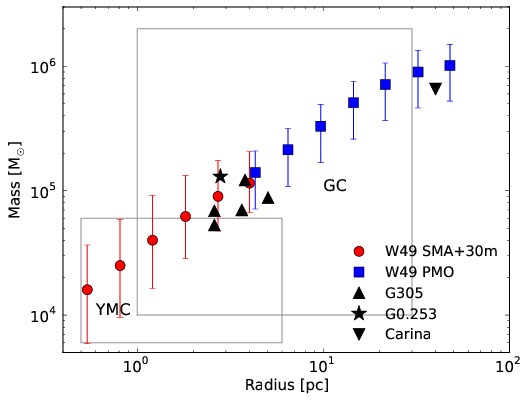}
\caption{
Mass vs radius for Galactic molecular clouds that may form (G0.253) or are indeed forming (the rest of plotted clouds) massive 
clusters ($M_\star>10^4$ M$_\odot$ in stellar mass). The filled red circles and blue squares show the total mass in the W49 GMC as a 
function of radius around the Welch ring from the observations presented in this paper. 
The black symbols are measurements compiled from the literature (only one value of mass and radius is given): 
the Galactic Center cloud G0.253 \citep{Longmore12a}, the Carina complex GMC \citep{Preibisch12a}, and the most massive clumps in G305 
\citep{Hindson10}. 
The typical regimes of stellar mass and radius for Galactic young massive clusters (YMCs) and globular clusters (GCs) are marked 
with boxes \citep{PZ10}. Error bars include systematic uncertainties from element/isotopologue abundances and excitation temperature 
(see Appendix \ref{App-D}).}
\label{fig:MvsR}
\end{figure}

\clearpage

\makeatletter 
\renewcommand{\thefigure}{A\@arabic\c@figure}
\makeatother
\setcounter{figure}{0}
\begin{figure}[t]
\centering
\begin{tabular}{cc}
\includegraphics[angle=0,width=0.50\columnwidth]{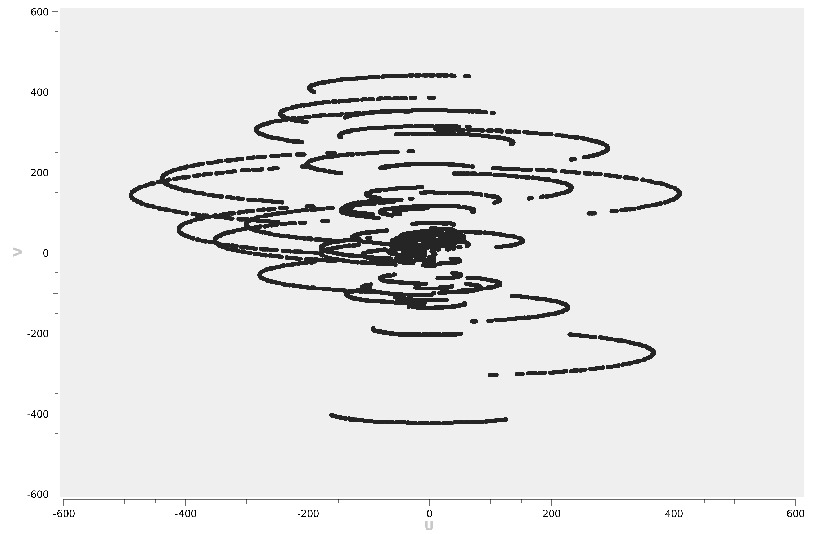} & 
\includegraphics[angle=0,width=0.45\columnwidth]{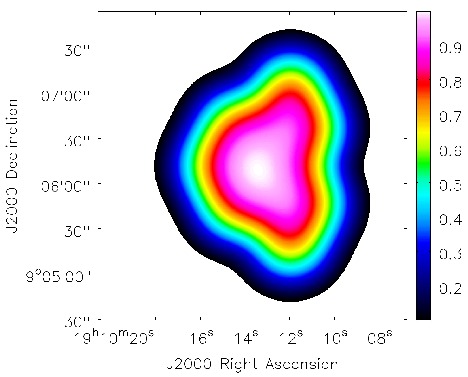} \\
\end{tabular}
\caption{{\it Left:} $u,v$ coverage in kilolambdas ($k\lambda$) of the concatenated SMA data set from the four array configurations. 
{\it Right:} Primary-beam response of the 11-pointing SMA mosaic. The images presented in this paper have been divided by this 
response to 
correct their flux scale.
 \label{fig:appa-sma}}
\end{figure}

\clearpage

\makeatletter 
\renewcommand{\thefigure}{B\@arabic\c@figure}
\makeatother
\setcounter{figure}{0}
\begin{figure}[t]
\centering
\includegraphics[angle=0,scale=0.4]{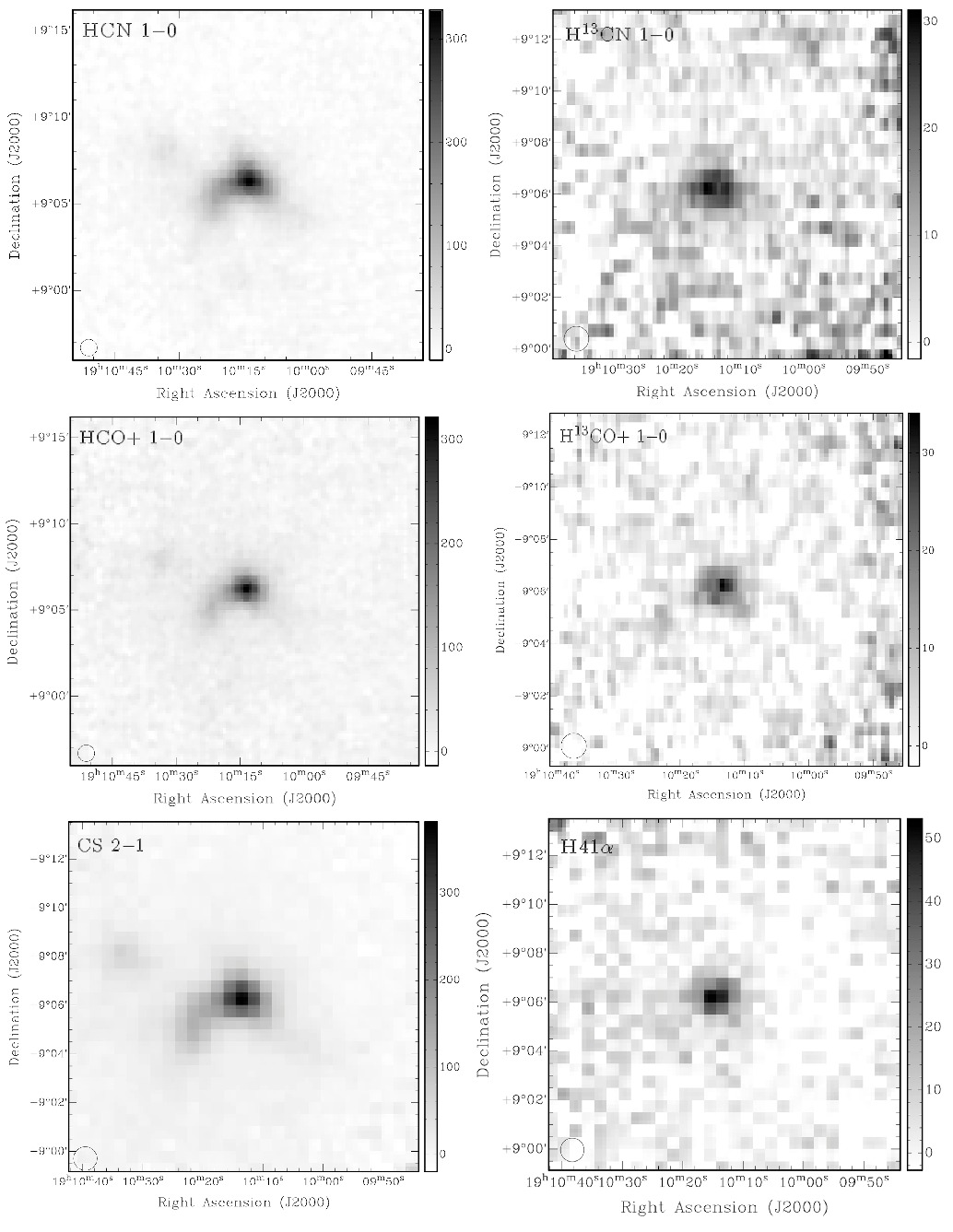}
\caption{PMO velocity-integrated (from $-12$ to 24 \kms) intensity maps of carbon molecules (beside CO and its 
isotopologues) with clear detections, plus the H41$\alpha$ recombination line. 
Units are K \kms~. The HPBW is 58\arcsec. 10\arcmin~ are equivalent to 32.3 pc.
The zoomed areas are different because the map sizes are different. 
The rms noises in the shown areas are  5 to 6 K \kms~ in all cases. 
\label{fig:W49-moms0-carbon-PMO}
}
\end{figure}

\clearpage

\begin{figure}[t] 
\centering
\includegraphics[angle=0,scale=0.5]{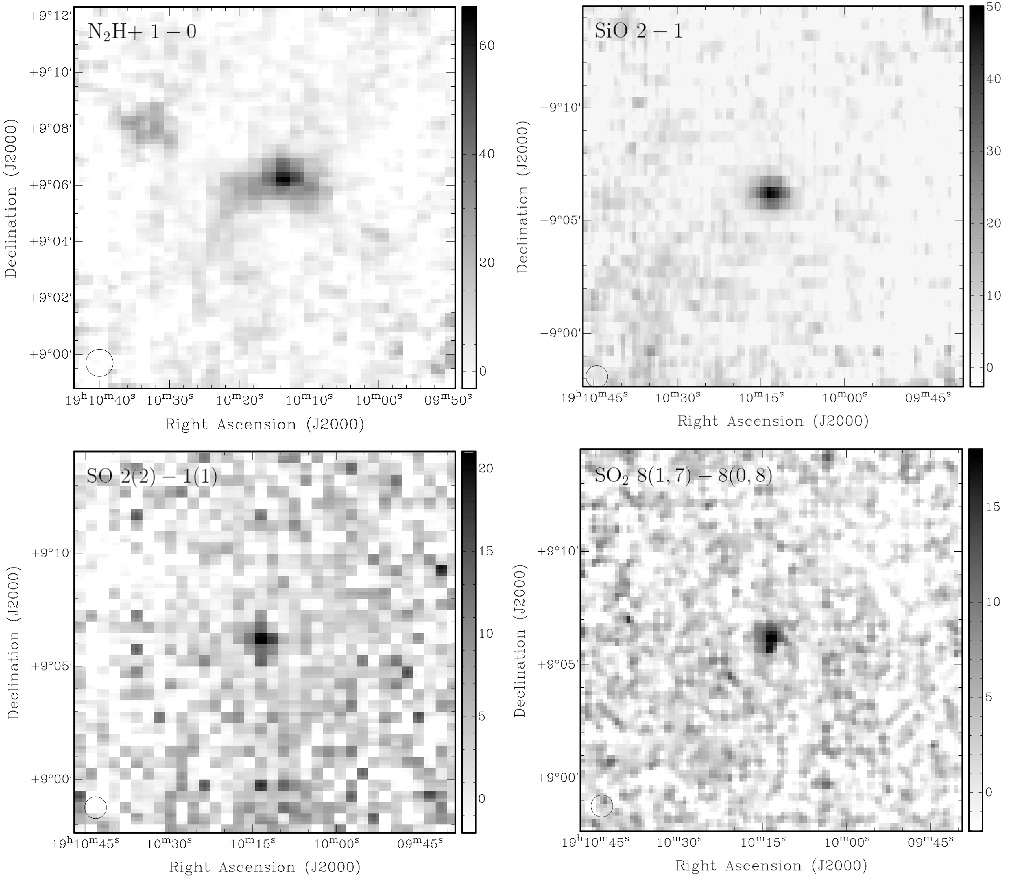}
\caption{PMO velocity-integrated (from $-12$ to 24 \kms) intensity maps of non-carbon molecules with clear detections. 
Units are K \kms~. The HPBW is 58\arcsec. 10\arcmin~ are equivalent to 32.3 pc.
The zoomed areas are different because the map sizes are different. 
The rms noises in the shown areas are 4 to 5 K \kms~ in all cases. 
\label{fig:W49-moms0-noncarbon-PMO}
}
\end{figure}

\clearpage 

\makeatletter 
\renewcommand{\thefigure}{C\@arabic\c@figure}
\makeatother
\setcounter{figure}{0}
\begin{figure}
\centering
\begin{tabular}{cc}
\includegraphics[angle=0,width=0.38\columnwidth]{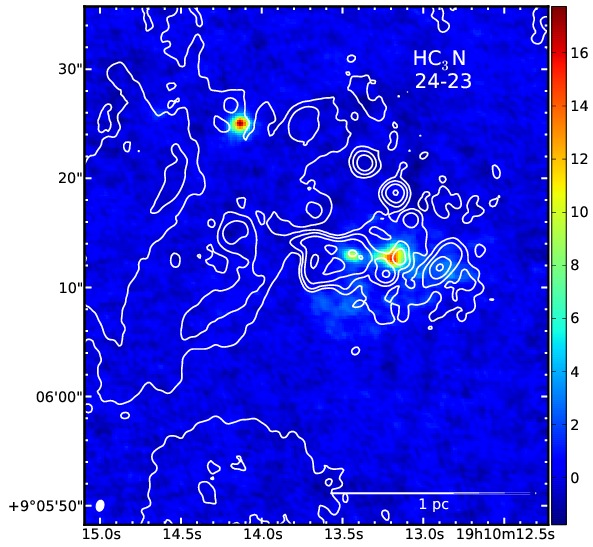} & 
\includegraphics[angle=0,width=0.38\columnwidth]{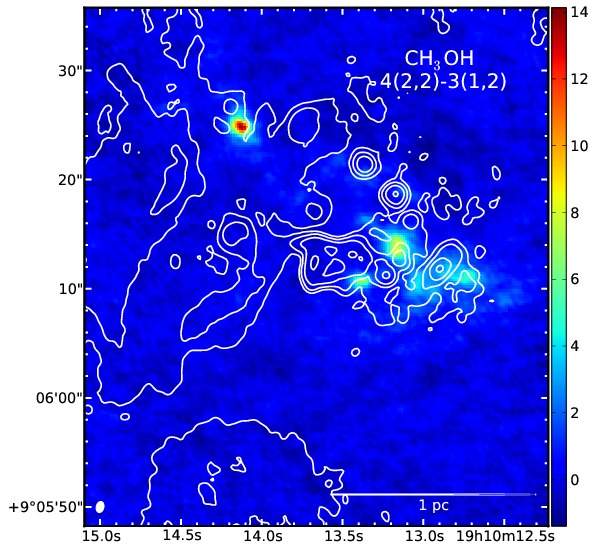} \\
\includegraphics[angle=0,width=0.38\columnwidth]{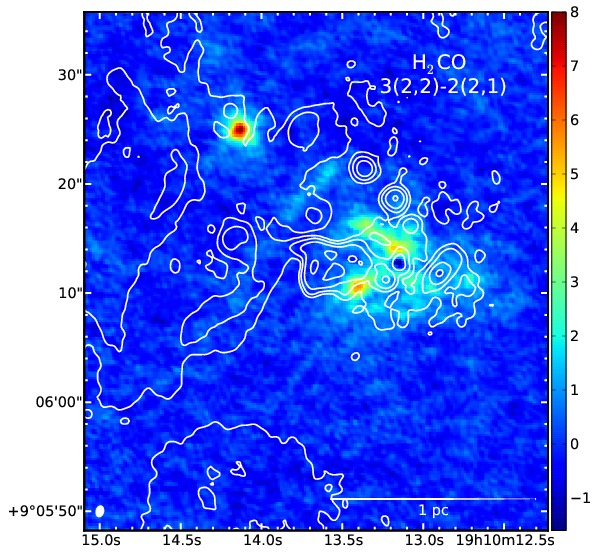} & 
\includegraphics[angle=0,width=0.38\columnwidth]{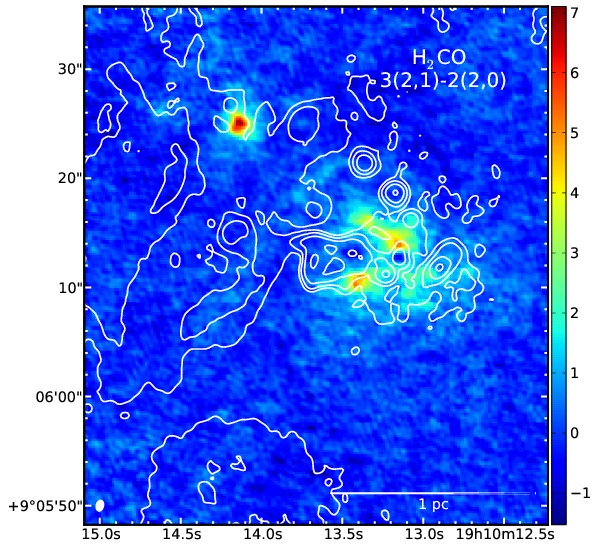} \\
\includegraphics[angle=0,width=0.38\columnwidth]{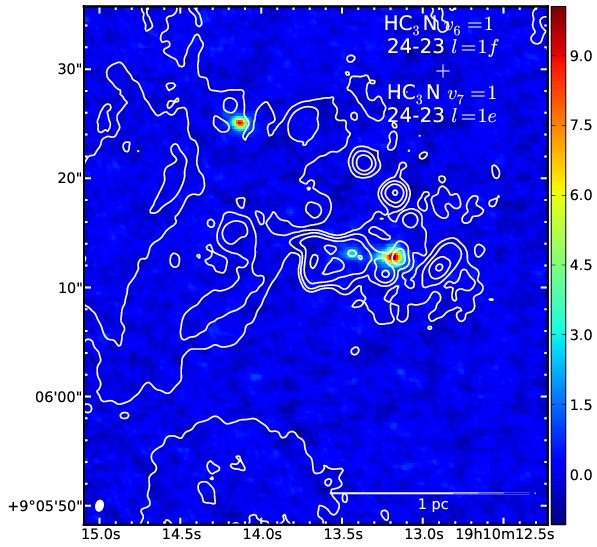} & 
\includegraphics[angle=0,width=0.38\columnwidth]{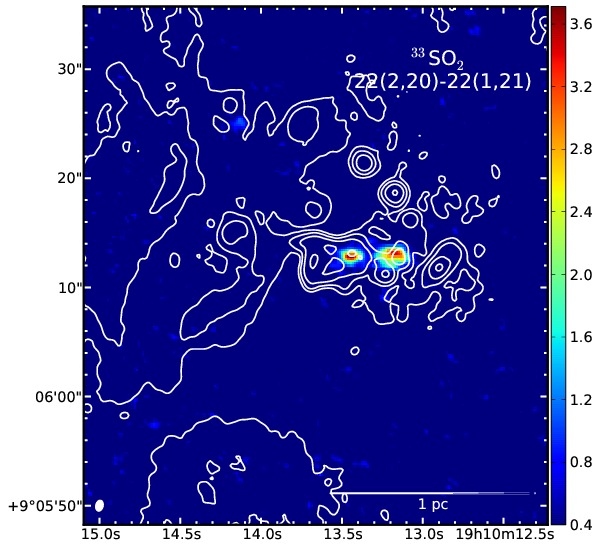} \\
\end{tabular}
\caption{(a) Velocity-integrated (moment 0) SMA mosaics, {\it a:} The color scale shows the velocity-integrated 
intensity in Jy beam$^{-1}$ 
km s$^{-1}$ using a linear stretching. 
The synthesized beam of these maps varies smoothly from $1.02\arcsec\times0.69\arcsec$, PA$=76.8^\circ$ at the lowest 
frequency (36.4 K per Jy beam$^{-1}$) to $0.98\arcsec\times0.64\arcsec$, PA$=74.1^\circ$ at the highest 
(31.5 K per Jy beam$^{-1}$).
White contours show the 3.6-cm free-free continuum at $-4,4,16,64$ and $256\times0.5$ 
mJy beam$^{-1}$, VLA HPBW=$0.80\arcsec\times0.78\arcsec$, PA=$-63.1^\circ$.}
 \label{fig:appb-smamos-1a}
\end{figure}

\clearpage

\setcounter{figure}{0}
\begin{figure}
\centering
\begin{tabular}{cc}
\includegraphics[angle=0,width=0.38\columnwidth]{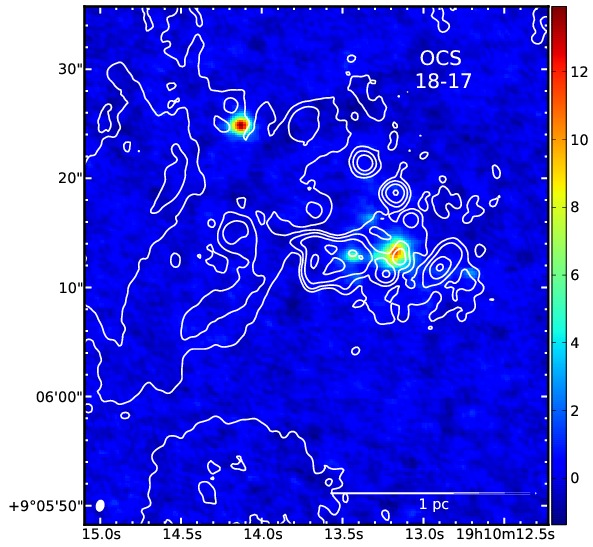} &
\includegraphics[angle=0,width=0.38\columnwidth]{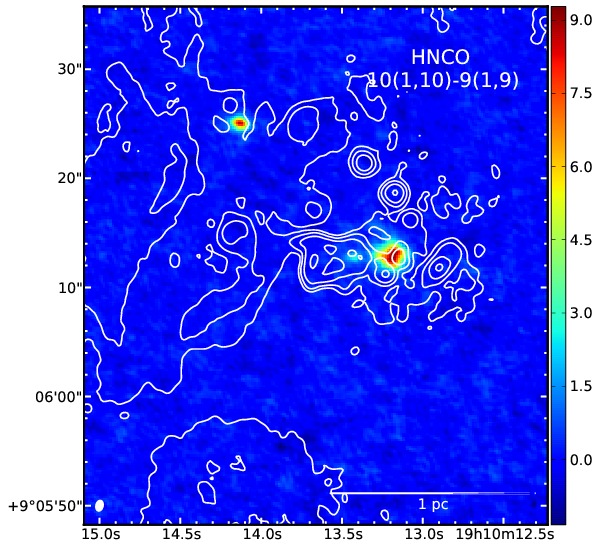} \\ 
\includegraphics[angle=0,width=0.38\columnwidth]{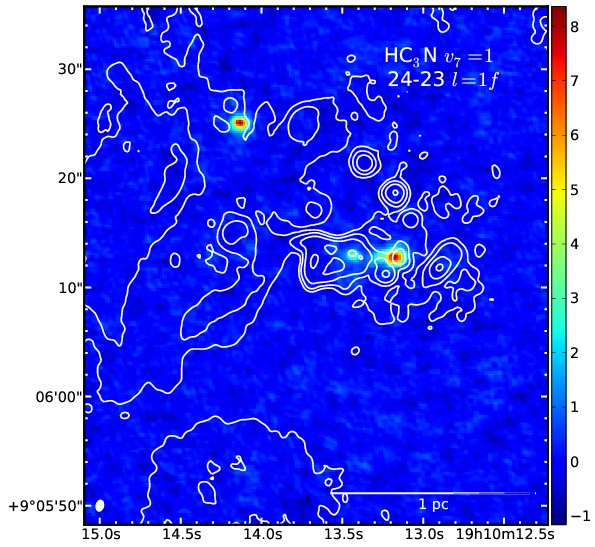} &
\includegraphics[angle=0,width=0.38\columnwidth]{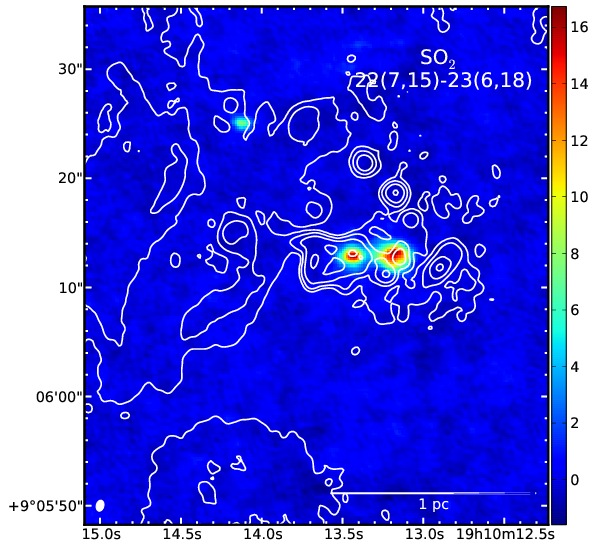} \\ 
\includegraphics[angle=0,width=0.38\columnwidth]{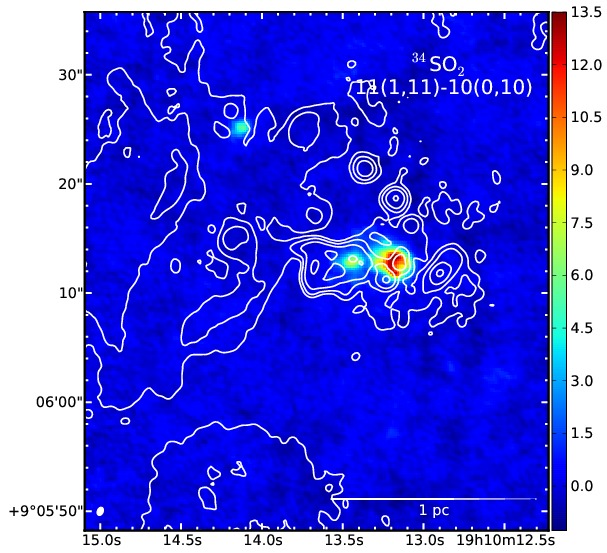} &
\includegraphics[angle=0,width=0.38\columnwidth]{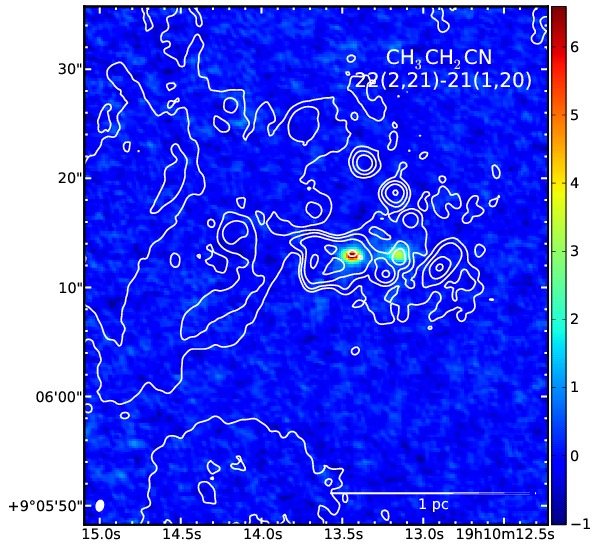} \\
\end{tabular}
\caption{(b) Velocity-integrated (moment 0) SMA mosaics, see {\it a} for description.}
 \label{fig:appb-smamos-1b}
\end{figure}

\clearpage

\setcounter{figure}{0}
\begin{figure}
\centering
\begin{tabular}{cc}
\includegraphics[angle=0,width=0.38\columnwidth]{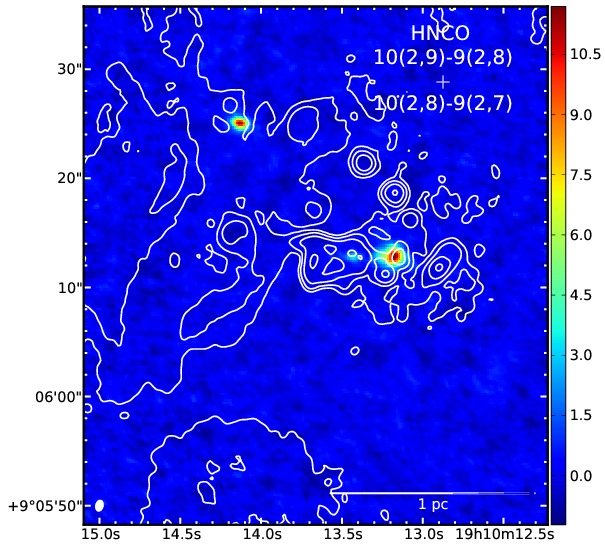} &
\includegraphics[angle=0,width=0.38\columnwidth]{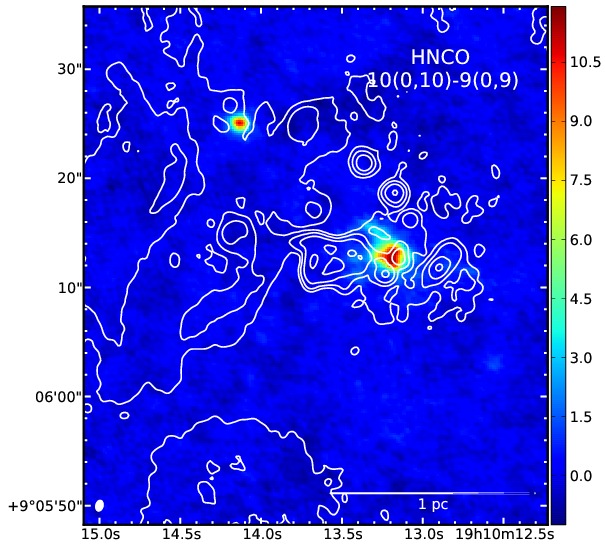} \\ 
\includegraphics[angle=0,width=0.38\columnwidth]{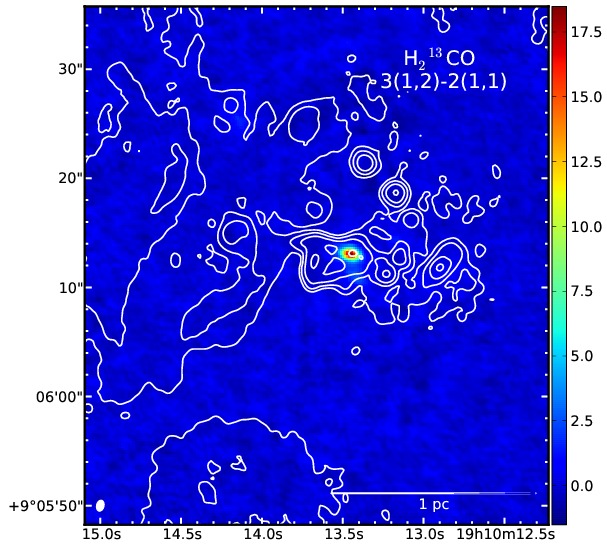} &
\includegraphics[angle=0,width=0.38\columnwidth]{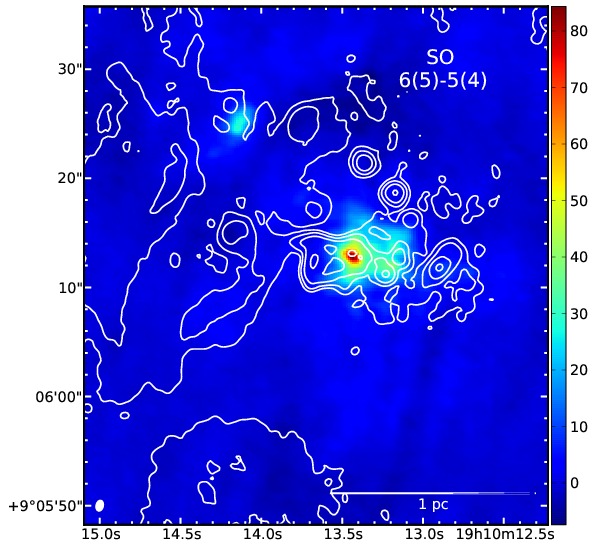} \\ 
\includegraphics[angle=0,width=0.38\columnwidth]{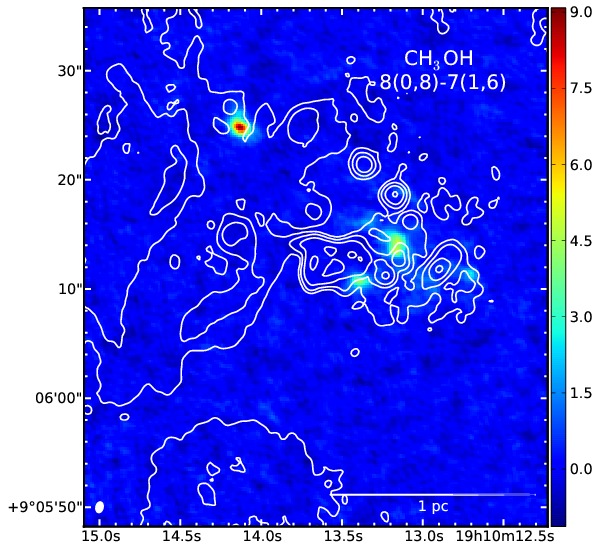} &
\includegraphics[angle=0,width=0.38\columnwidth]{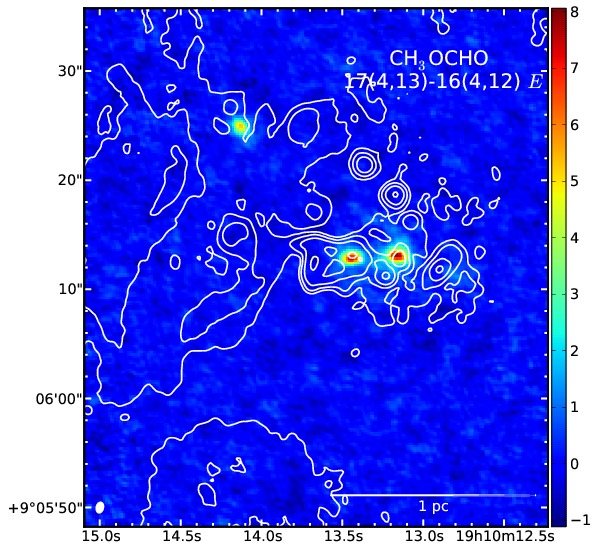} \\
\end{tabular}
\caption{(c) Velocity-integrated (moment 0) SMA mosaics, see {\it a} for description.}
 \label{fig:appb-smamos-1c}
\end{figure}

\clearpage

\setcounter{figure}{0}
\begin{figure}
\centering
\begin{tabular}{cc}
\includegraphics[angle=0,width=0.38\columnwidth]{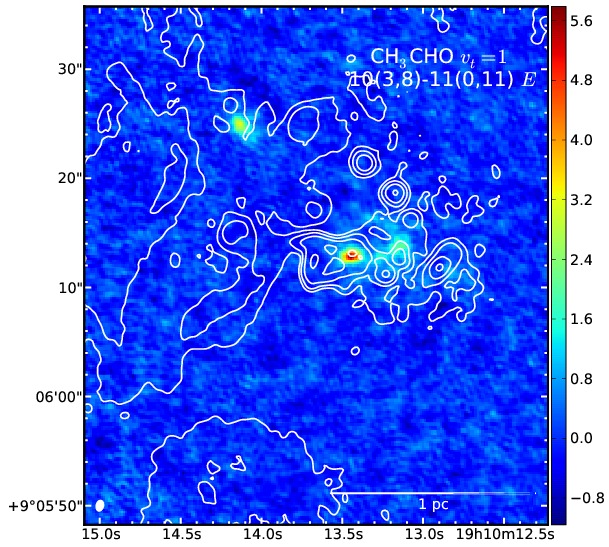} &
\includegraphics[angle=0,width=0.38\columnwidth]{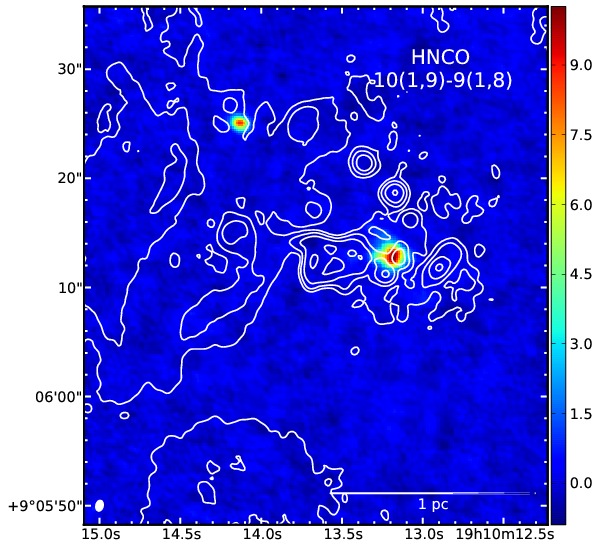} \\ 
\includegraphics[angle=0,width=0.38\columnwidth]{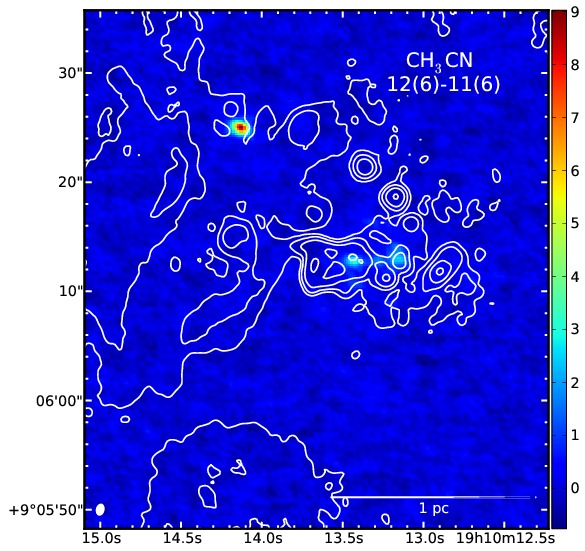} &
\includegraphics[angle=0,width=0.38\columnwidth]{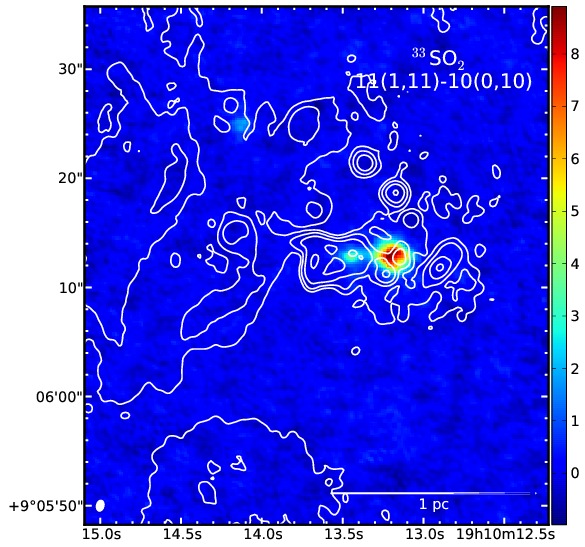} \\ 
\includegraphics[angle=0,width=0.38\columnwidth]{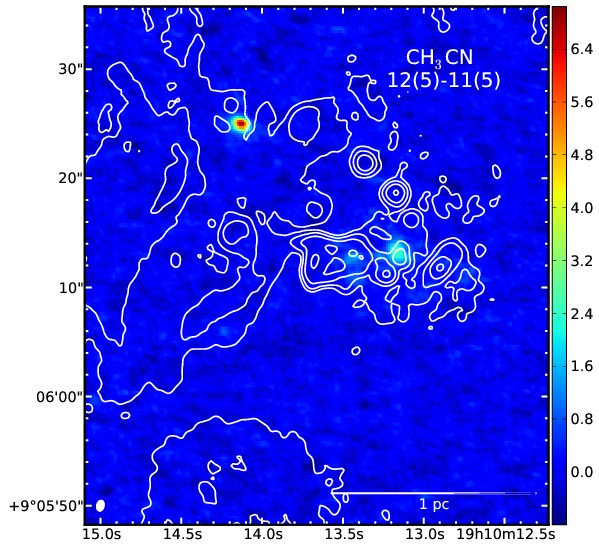} &
\includegraphics[angle=0,width=0.38\columnwidth]{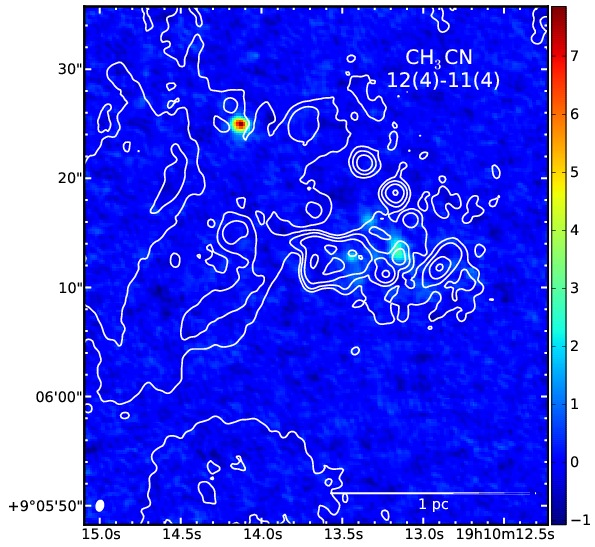} \\
\end{tabular}
\caption{(d) Velocity-integrated (moment 0) SMA mosaics, see {\it a} for description.}
 \label{fig:appb-smamos-1d}
\end{figure}

\clearpage

\setcounter{figure}{0}
\begin{figure}
\centering
\begin{tabular}{cc}
\includegraphics[angle=0,width=0.38\columnwidth]{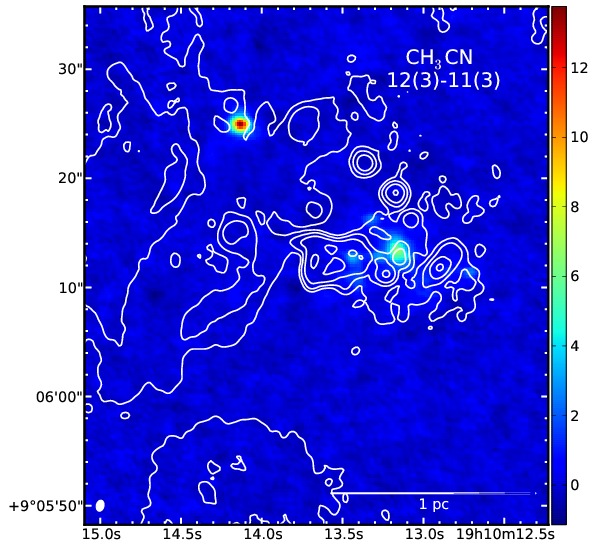} &
\includegraphics[angle=0,width=0.38\columnwidth]{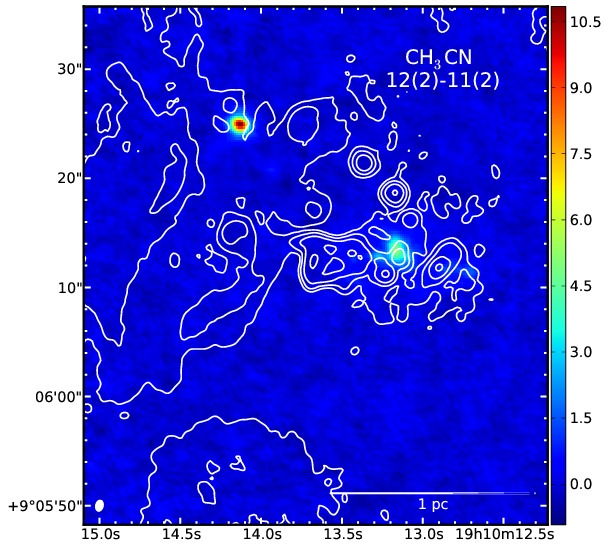} \\ 
\includegraphics[angle=0,width=0.38\columnwidth]{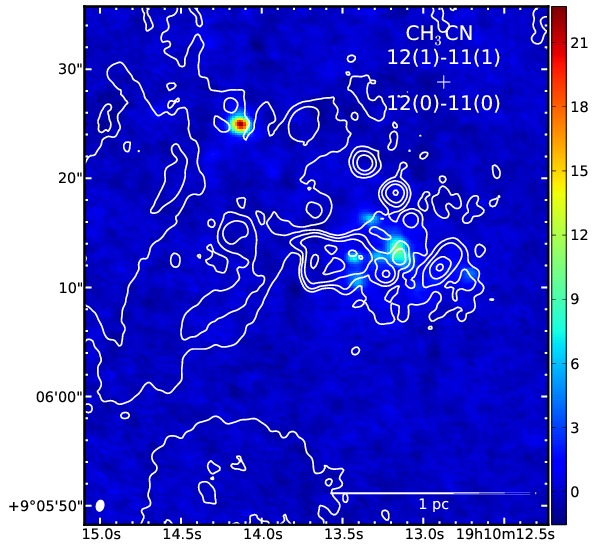} &
\includegraphics[angle=0,width=0.38\columnwidth]{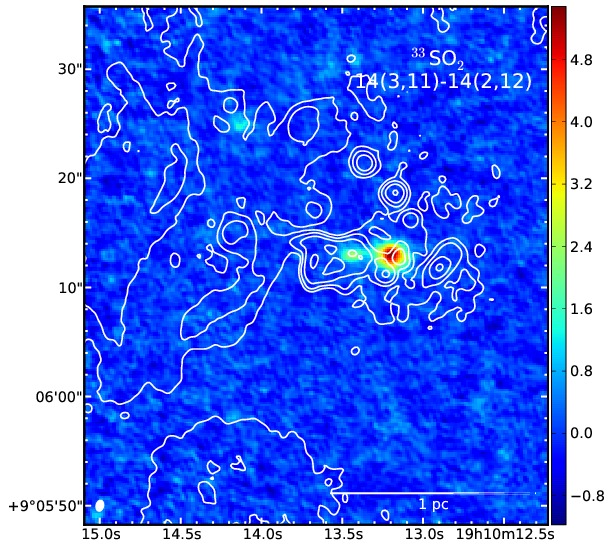} \\ 
\includegraphics[angle=0,width=0.38\columnwidth]{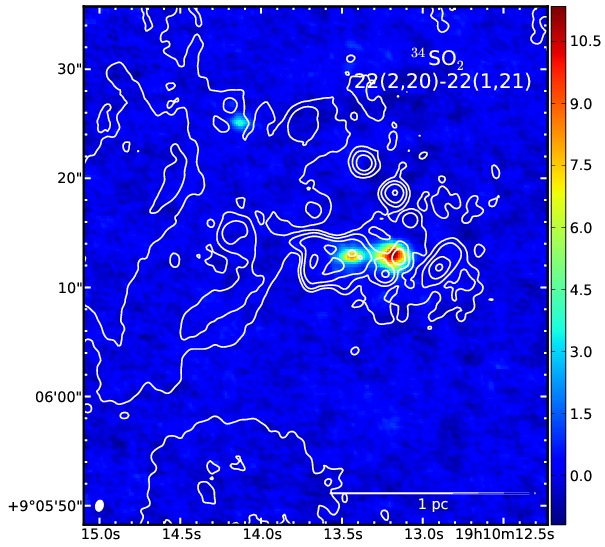} &
\includegraphics[angle=0,width=0.38\columnwidth]{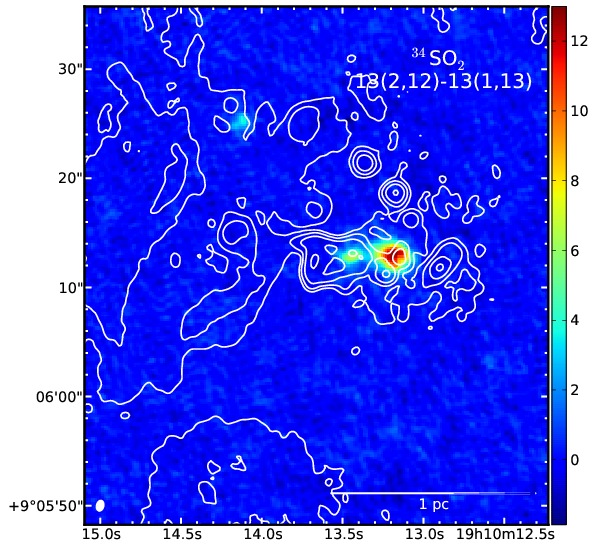} \\
\end{tabular}
\caption{(e) Velocity-integrated (moment 0) SMA mosaics, see {\it a} for description.}
 \label{fig:appb-smamos-1e}
\end{figure}

\clearpage

\setcounter{figure}{0}
\begin{figure}
\centering
\begin{tabular}{cc}
\includegraphics[angle=0,width=0.38\columnwidth]{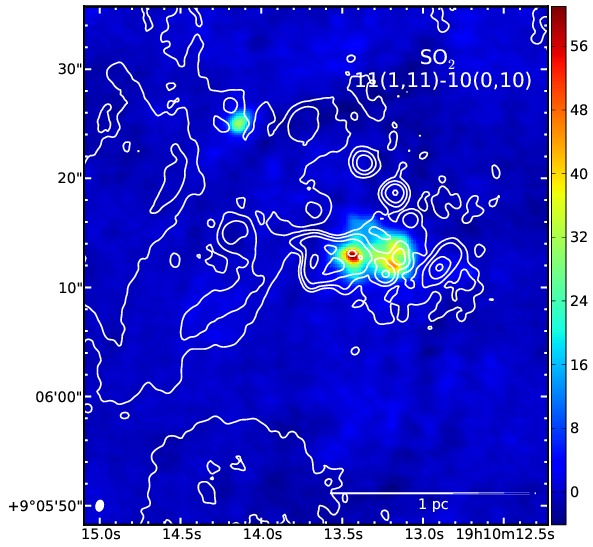} &
\includegraphics[angle=0,width=0.38\columnwidth]{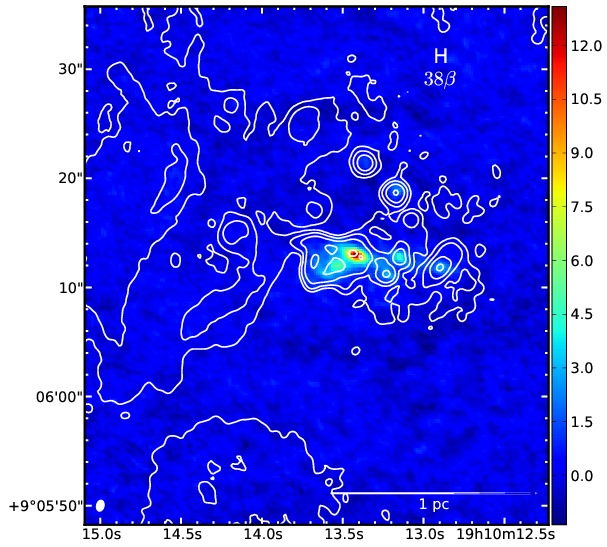} \\ 
\includegraphics[angle=0,width=0.38\columnwidth]{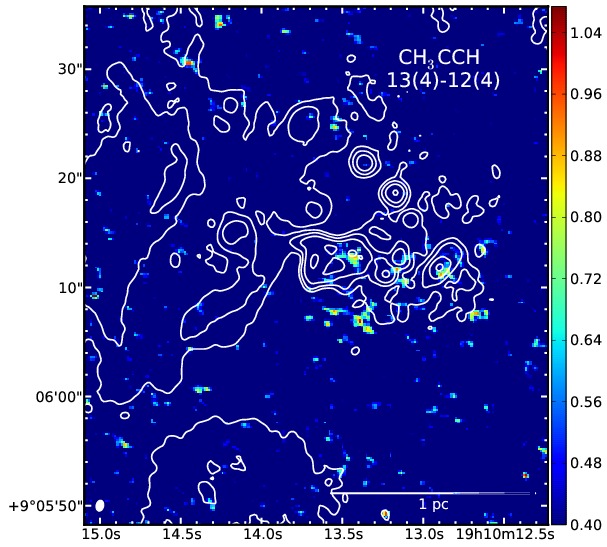} &
\includegraphics[angle=0,width=0.38\columnwidth]{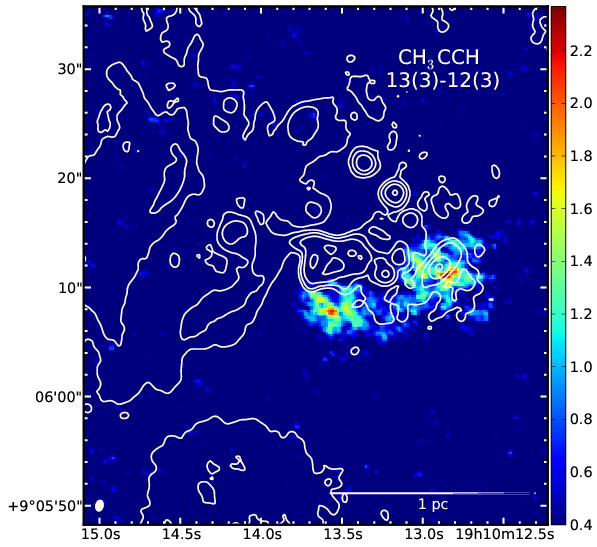} \\ 
\includegraphics[angle=0,width=0.38\columnwidth]{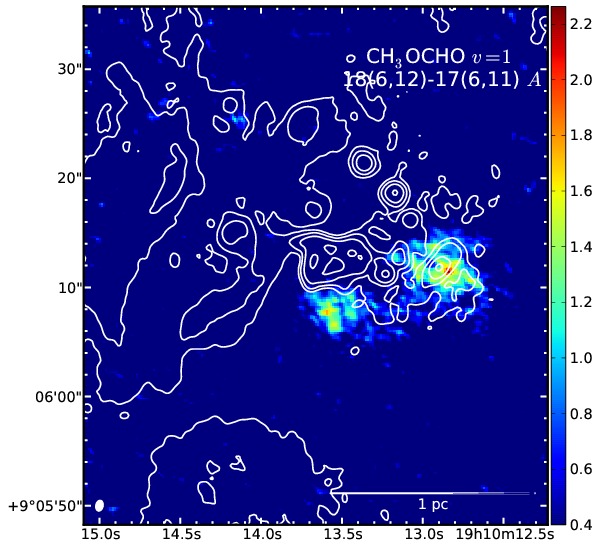} &
\includegraphics[angle=0,width=0.38\columnwidth]{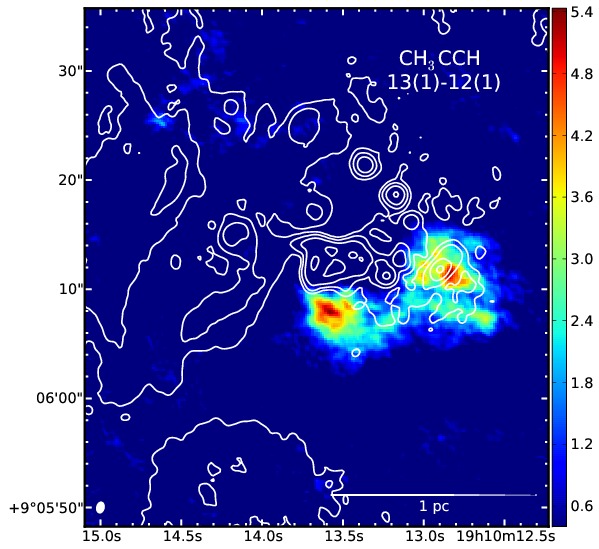} \\
\end{tabular}
\caption{(f) Velocity-integrated (moment 0) SMA mosaics, see {\it a} for description.}
 \label{fig:appb-smamos-1f}
\end{figure}

\clearpage

\setcounter{figure}{0}
\begin{figure}
\centering
\begin{tabular}{cc}
\includegraphics[angle=0,width=0.38\columnwidth]{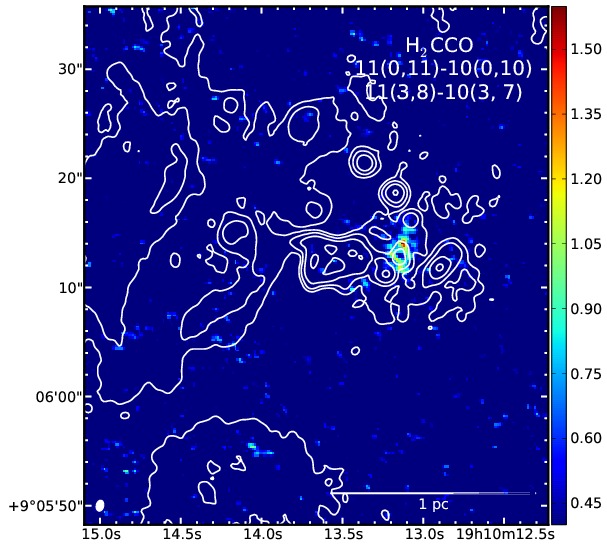} &
\includegraphics[angle=0,width=0.38\columnwidth]{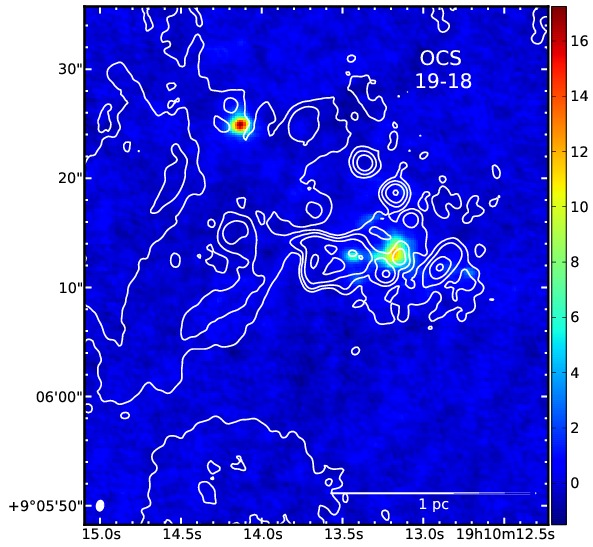} \\ 
\includegraphics[angle=0,width=0.38\columnwidth]{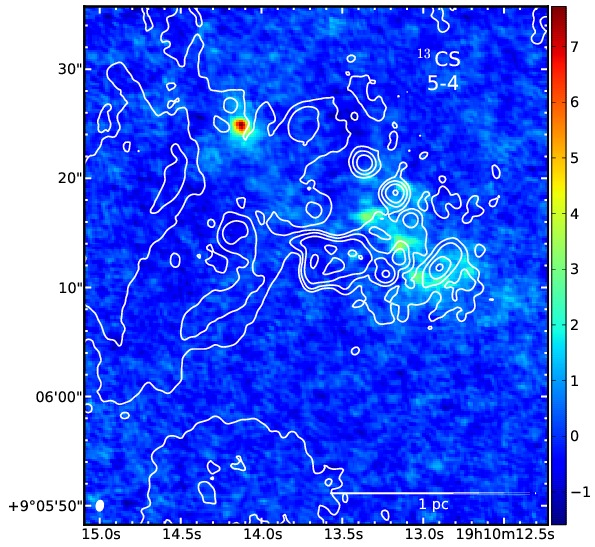} &
\includegraphics[angle=0,width=0.38\columnwidth]{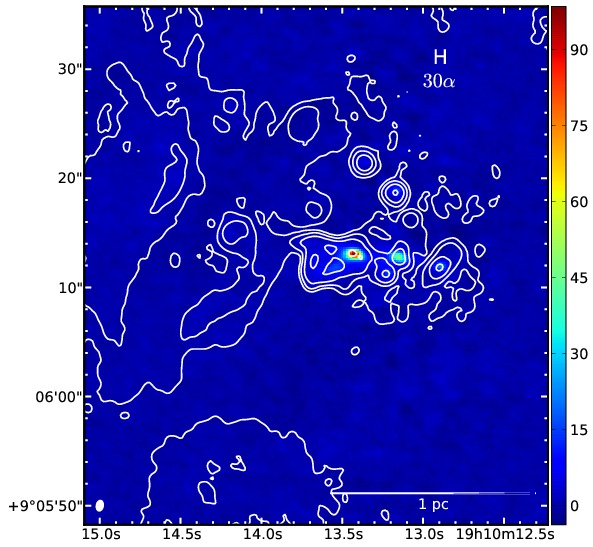} \\ 
\includegraphics[angle=0,width=0.38\columnwidth]{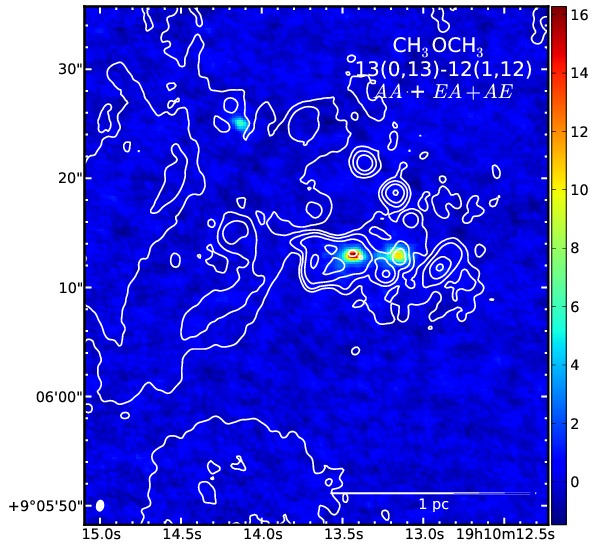} &
\includegraphics[angle=0,width=0.38\columnwidth]{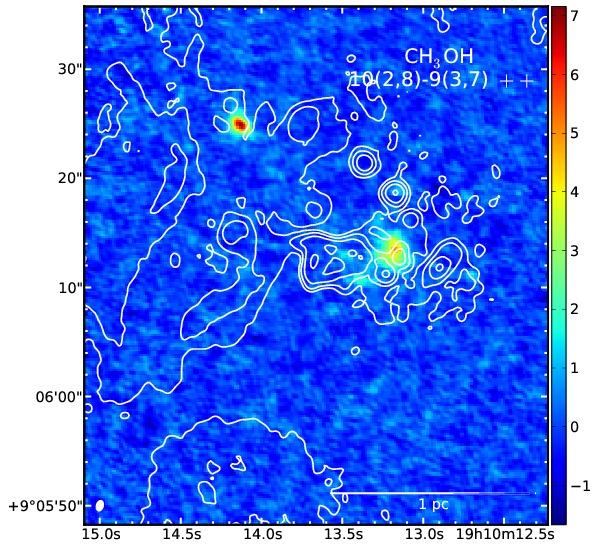} \\
\end{tabular}
\caption{(g) Velocity-integrated (moment 0) SMA mosaics, see {\it a} for description.}
 \label{fig:appb-smamos-1g}
\end{figure}

\clearpage

\setcounter{figure}{0}
\begin{figure}
\centering
\begin{tabular}{cc}
\includegraphics[angle=0,width=0.38\columnwidth]{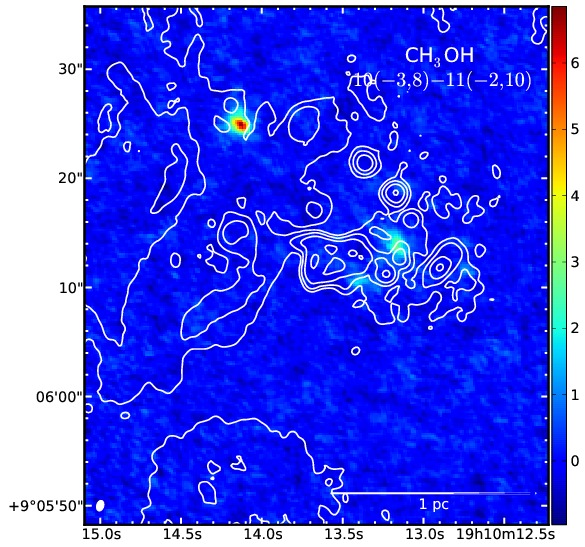} &
\includegraphics[angle=0,width=0.38\columnwidth]{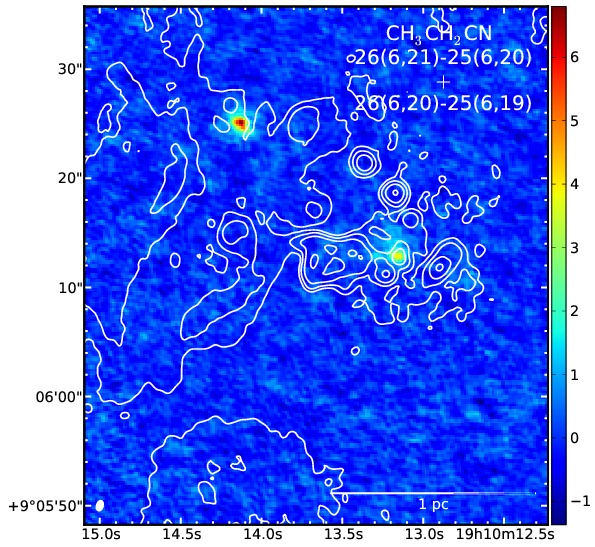} \\ 
\includegraphics[angle=0,width=0.38\columnwidth]{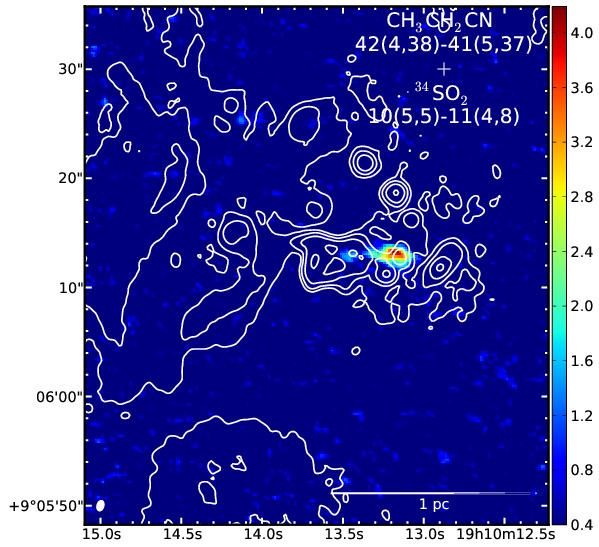} &
\includegraphics[angle=0,width=0.38\columnwidth]{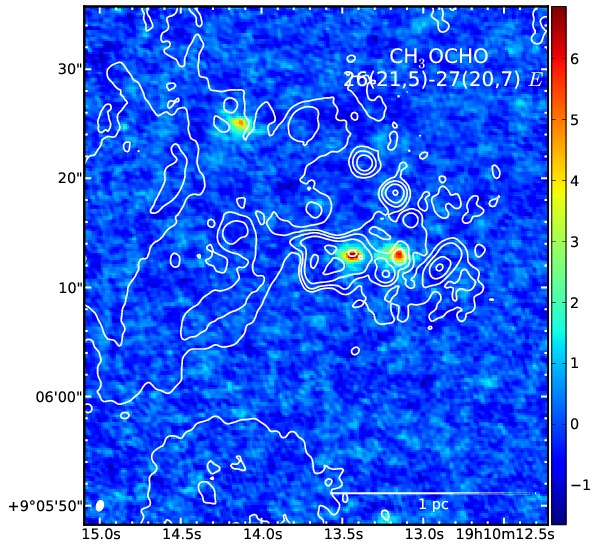} \\ 
\includegraphics[angle=0,width=0.38\columnwidth]{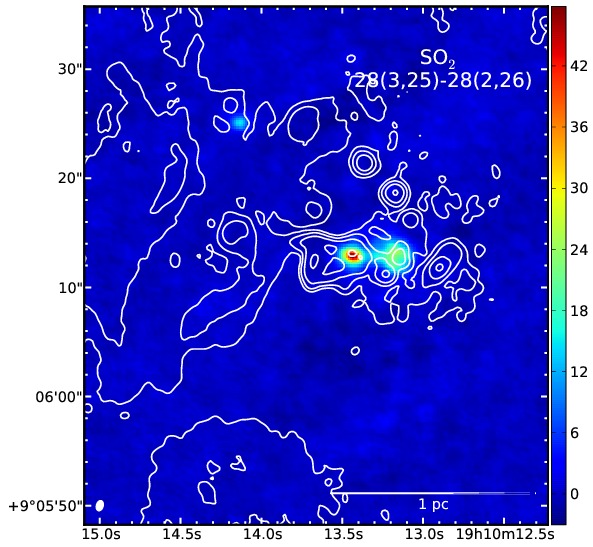} \\
\end{tabular}
\caption{(h) Velocity-integrated (moment 0) SMA mosaics, see {\it a} for description.}
 \label{fig:appb-smamos-1h}
\end{figure}

\clearpage

\begin{deluxetable}{cccccc} 
\tabletypesize{\scriptsize} 
\tablecolumns{6} 
\tablewidth{0pt}
\tablecaption{SMA observations \label{tab:SMA-obs}} 
\tablehead{
\colhead{Array} & \colhead{Epoch} & \colhead{Flux} & \colhead{Phase} & \colhead{Bandpass} & \colhead{$\tau_\mathrm{225GHz}$} \\
\colhead{} & \colhead{} & \colhead{Calibrator} & \colhead{Calibrator} & \colhead{Calibrator} &
}
\startdata
Subcompact & 15 Aug 2010 & Callisto & 1751+096 & 3C454.3 & 0.05 to 0.08 \\
Compact  & 19 Jun 2010 & Neptune & 1751+096 & 3C279 & 0.08 \\
Extended & 10 Sep 2010 & Callisto & 1751+096 & 3C454.3 & 0.08  \\
Very extended & 15 Jul 2010 & Callisto & 1751+096 & 3C454.3 & 0.08
\enddata
\end{deluxetable}

\begin{deluxetable}{ccccc}
\tabletypesize{\scriptsize} 
\tablecolumns{5} 
\tablewidth{0pt}
\tablecaption{Lines detected with PMO\tablenotemark{a}\label{tab:W49-lines-PMO}} 
\tablehead{
\colhead{Molecule} & \colhead{Transition} & \colhead{$\nu_0$} & $E_\mathrm{U}$ & Integrated flux  \\
\colhead{} & \colhead{} & \colhead{[GHz]} & \colhead{[K]} & \colhead{[Jy km s$^{-1}$]}
}
\startdata
$^{12}$CO & $1-0$ & 115.2712 &  5.53 &  $5.005268\times10^6$  \\ 
$^{13}$CO & $1-0$ & 110.2013 &  5.29 &  $7.35425\times10^5$  \\ 
C$^{18}$O & $1-0$ & 109.7821 & 5.27 &   $3.0754\times10^4$  \\
\cline{1-5} \\
HCN & $1-0$ & 88.6316 & 4.25 &  $ 5.5959\times10^4$\tablenotemark{b}   \\
H$^{13}$CN & $1-0$ & 86.3399 & 4.14 & $1.445\times10^3$  \\
HCO+ & $1-0$ & 89.1885 & 4.28 & $5.4033\times10^4$\tablenotemark{b}   \\
H$^{13}$CO+ & $1-0$ & 86.7542 &  4.16  & $1.121\times10^3$   \\
CS & $2-1$ & 97.9809 & 7.05 & $5.4608\times10^4$\tablenotemark{b}   \\ 
H & $41\alpha$ & 92.0344 & -- & $1.262\times10^3$  \\ 
\cline{1-5} \\ 
N$_2$H+ & $1-0$ & 93.1734 & 4.47 & $6.273\times10^3$ \\ 
SiO & $2-1$ & 86.8469 & 6.25 & $1.814\times10^3$  \\ 
SO & $2(2)-1(1)$ & 86.0939 & 19.31 & $7.81\times10^2$  \\ 
SO$_2$ & $8(1,7)-8(0,8)$ & 83.6880 & 36.71 & $5.25\times10^2$  
\enddata
\tablenotetext{a}{~Lines detected with the PMO 14m telescope. The first four columns indicate, respectively, the molecule 
or atom, the transition, the rest frequency, and the upper-level energy of the transition. The fifth column lists the 
velocity-integrated flux of the maps in Jy beam$^{-1}$ km s$^{-1}$. The intensity conversion factor is 0.027 K per Jy beam$^{-1}$ at 
the rest frequency of the CO 1--0. The noise in the  moment 0 maps ranges from 2 to 10 K km s$^{-1}$. 
}
\tablenotetext{b}{~From the $^{12}$C to $^{13}$C line ratios, the average optical depths of the main isotopologues 
are $\tau_\mathrm{HCN}=1.4$, $\tau_\mathrm{HCO+}=0.8$, and $\tau_\mathrm{CS}<0.1$.}
\end{deluxetable}

\begin{deluxetable}{ccccccccccc}
\tabletypesize{\scriptsize} 
\tablecolumns{11} 
\tablewidth{0pt}
\tablecaption{Line Fitting at Peak Position of PMO data\tablenotemark{a}\label{tab:W49-specfit-PMO}} 
\tablehead{
\colhead{Molecule} & \colhead{Transition} &  \multicolumn{3}{c}{Component 1} 
& \multicolumn{3}{c}{Component 2} & \multicolumn{3}{c}{Component 3} \\
\cline{3-5} \cline{6-8} \cline{9-11} \\
\colhead{or Atom} & \colhead{}  &\colhead{$V$} & \colhead{$A$} & \colhead{FWHM} 
& \colhead{$V$} & \colhead{$A$} & \colhead{FWHM} 
& \colhead{$V$} & \colhead{$A$} & \colhead{FWHM} \\
\colhead{} & \colhead{}  & \colhead{[km/s]} & \colhead{[K]} & \colhead{[km/s]} 
& \colhead{[km/s]} & \colhead{[K]} & \colhead{[km/s]} 
& \colhead{[km/s]} & \colhead{[K]} & \colhead{[km/s]}
}
\startdata
$^{12}$CO & $1-0$ &  -- & -- & -- & -- & -- & -- & -- & -- & --  \\ 
$^{13}$CO & $1-0$ &  $3.9\pm0.2$ & $9.3\pm0.2$ & $8.5\pm0.5$ & $12.0\pm0.2$ & $11.8\pm0.3$ & $5.9\pm0.6$ & $16.9\pm0.5$ & $1.7\pm0.5$ & $4.1\pm0.9$ \\ 
C$^{18}$O & $1-0$ &  $4.0\pm0.4$ & $0.9\pm0.1$ & $8.7\pm0.9$ & $12.3\pm0.3$ & $1.3\pm0.2$ & 
$6.3\pm0.7$ & $16.4\pm0.3$ & $0.3\pm0.2$ & $1.0\pm0.8$ \\
\cline{1-11} \\
HCN & $1-0$ &  $2.5\pm0.3$ & $3.4\pm0.2$ & $13.8\pm2.5$ & $11.8\pm0.3$ & $ 2.3\pm0.4$ & $4.0\pm1.8$ & 
$18.0\pm0.9$ & $0.8\pm0.2$ & $10.9\pm3.7$ \\
H$^{13}$CN & $1-0$ &  $ -2.2\pm0.9$ & $0.3\pm0.2$ & $8.3\pm3.4$ & $7.6\pm0.9$ & $0.3\pm0.1$ & $9.3\pm3.5$ & -- & -- & -- \\
HCO+ & $1-0$ &  $2.0\pm0.3$ & $5.1\pm0.3$ & $7.7\pm2.0$ &  $10.8\pm0.3$ & $3.6\pm0.3$ & $4.8\pm1.8$ 
& -- & -- & -- \\
H$^{13}$CO+ & $1-0$ &  $2.9\pm0.5$ & $0.3\pm0.2$ & $5.1\pm2.7$ & $10.3\pm1.0$ & $0.4\pm0.1$ & 
$13.2\pm3.9$\tablenotemark{b} & -- & -- & -- \\
CS & $2-1$ & $3.2\pm0.2$ & $3.6\pm0.2$ & $7.9\pm1.9$ & $11.7\pm0.2$ & $4.8\pm0.2$ & $7.8\pm1.7$ 
& -- & -- & --\\ 
H & $41\alpha$ &  $12.2\pm0.6$ & $0.4\pm0.1$ & $30.2\pm4.2$ & -- & -- & -- & -- & -- & -- \\ 
\cline{1-11} \\ 
N$_2$H+ & $1-0$ &  $3.0\pm0.7$ & $0.4\pm0.2$ & $8.6\pm2.9$ & $12.6\pm0.7$ & $0.6\pm0.1$ & $10.6\pm3.0$ 
& -- & -- & --\\ 
SiO & $2-1$ &  $6.9\pm0.4$ & $0.7\pm0.1$ & $12.8\pm2.6$ & -- & -- & -- & -- & -- & -- \\ 
SO & $2(2)-1(1)$ &  $9.1\pm0.7$ & $0.4\pm0.2$ & $10.7\pm3.4$ & -- & -- & -- & -- & -- & -- \\ 
SO$_2$ & $8(1,7)-8(0,8)$ & $8.4\pm0.8$ & $0.3\pm0.2$ & $13.6\pm3.9$ & -- & -- & -- & -- & -- & --  
\enddata
\tablenotetext{a}{Line parameters from Gaussian fitting at the peak of integrated emission in the PMO maps: 
$\alpha\mathrm{(J2000)}=19^{\mathrm{h}}~10^{\mathrm{m}}~14.2^{\mathrm{s}},
~\delta\mathrm{(J2000)}=9^\circ~6\arcmin~23\arcsec$. The first two columns indicate, respectively, the molecule 
or atom and the transition. The $^{12}$CO spectrum 
is too complex to be fitted by 3 or less Gaussians. Most of the lines are better described by two Gaussian components, 
whereas a few of them are single-peaked (e.g., H$41\alpha$).}
\tablenotetext{b}{The second peak of H$^{13}$CO+ 1--0 appears quite wide, possibly due to a baseline difference at both 
extremes of the line emission.}
\end{deluxetable}

\begin{deluxetable}{cccc}
\tabletypesize{\scriptsize} 
\tablecolumns{4} 
\tablewidth{0pt}
\tablecaption{Mass structure of W49N from SMA+BOLOCAM}\tablenotemark{a}\label{tab:flux-sma} 
\tablehead{
\colhead{Intensity threshold} & \colhead{Surface density threshold} & \colhead{Flux from dust} & H$_2$ mass  \\
\colhead{$I>$ [Jy beam$^{-1}$]} & \colhead{$\Sigma>$ [g cm$^{-2}$]} & \colhead{[Jy]} & \colhead{[M$_\odot$]}    \\
}
\startdata
0.025 & $2.1\pm1.4$    &  17.017 &  $1.484\times10^5 \pm 9.89\times10^4$ \\
0.050 & $4.0\pm2.7$    &  12.745 &  $1.111\times10^5 \pm 7.41\times10^4$ \\
0.100 & $8.1\pm5.4$   &  9.216  &  $8.03\times10^4 \pm 5.36\times10^4$ \\
0.200 & $16.3\pm10.8$   &  7.462  &  $6.51\times10^4 \pm 4.33\times10^4$ \\
0.400 & $32.6\pm21.6$  &  6.267  &  $5.47\times10^4 \pm 3.64\times10^4$ \\
0.800 & $65.0\pm43.3$ &  4.253  &  $3.71\times10^4 \pm 2.47\times10^4$ \\
1.600 & $130.0\pm86.7$ &  2.239  &  $1.95\times10^4 \pm 1.30\times10^4$
\enddata
\tablenotetext{a}{The masses and threshold mass surface densities correspond to $\kappa_\mathrm{1mm}=0.25$
cm$^2$ g$^{-1}$  (per unit mass of dust). 
The quoted range of values correspond to variations in $\kappa_\mathrm{1mm}=0.25$ from 
0.5 to 0.1 cm$^2$ g$^{-1}$.  
}
\end{deluxetable}

\begin{center} 
\begin{deluxetable}{cccccc}   
\tabletypesize{\scriptsize}
\tablecolumns{6} 
\tablecaption{Molecular lines detected with SMA\tablenotemark{a} \label{tab:sma-lines}}
\tablewidth{0pt}
\tablehead{
\colhead{Species} & \colhead{Transition} & \colhead{$\nu_0$} [GHz] & \colhead{$E_u$} [K] & Integrated flux [Jy km s$^{-1}$] & Note
}   
\startdata
\multicolumn{6}{c}{LSB Low} \\ 
\cline{1-6}
HC$_3$N	& 24--23 &  218.3247 & 130.9 & 322 & \\
CH$_3$OH & 4(2,2)--3(1,2) & 218.4400 & 45.4 & 372 & \\
H$_2$CO & 3(2,2)--2(2,1) &  218.4756 & 68.0 & 282 & \\
-- & 3(2,1)--2(2,0) & 218.7600 & 68.1 & 294 & \\
HC$_3$N $v_6=1$ & 24--23 $l=1f$ & 218.8543 & 848.9 & 63 & (1)\\
HC$_3$N $v_7=1$ & 24--23 $l=1e$ & 218.8608 & 452.1 & 63 & (1) \\
$^{33}$SO$_2$ & 22(2,20)--22(1,21) & 218.8754 & 251.7 & 38 & \\
OCS & 18--17 & 218.9033 &  99.8 & 197 & \\ 
HNCO & 10(1,10)--9(1,9) & 218.9810 & 101.0 & 89 & \\
HC$_3$N $v_7=1$ & 24--23 $l=1f$ & 219.1737 & 452.3 & 60 & \\
SO$_2$ & 22(7,15)--23(6,18) & 219.2759 & 352.7 & 203 & \\
$^{34}$SO$_2$ & 11(1,11)--10(0,10) & 219.3550 & 60.1 & 305 & \\
CH$_3$CH$_2$CN & 22(2,21)--21(1,20) & 219.4636 & 112.4 & 33 & \\
C$^{18}$O & 2--1 & 219.5603  & 15.8 & $1.6943\times10^4$ & \\
HNCO & 10(2,9)--9(2,8) & 219.7338  & 228.4 & 80 & (1) \\
-- & 10(2,8)--9(2,7) & 219.7371  & 228.2 & 80 & (1) \\
-- & 10(0,10)--9(0,9) & 219.7982  & 58.0 & 198 & \\
H$_2$$^{13}$CO & 3(1,2)--2(1,1) & 219.9084 & 32.9 & 63 & \\
SO & 6(5)--5(4) & 219.9494  & 34.9 & $3.042\times10^3$ & \\
CH$_3$OH & 8(0,8)--7(1,6) & 220.0784 & 96.6 & 118 & \\
CH$_3$OCHO & 17(4,13)--16(4,12) $E$ & 220.1668 & 103.1 & 93 & \\
CH$_3$CHO $v_t =1$ & 10(3,8)--11(0,11) $E$ &  220.1800 & 275.4 & 63 & \\
\cline{1-6}
\multicolumn{6}{c}{LSB High} \\ 
\cline{1-6}
$^{13}$CO & 2--1 & 220.3986 & 15.86 & $1.23553\times10^5$ & \\
HNCO & 10(1,9)--9(1,8) & 220.5847 &  101.5 & 90 & \\
CH$_3$CN & 12(6)-11(6) & 220.5944  & 325.8  & 53 & \\
$^{33}$SO$_2$ & 11(1,11)--10(0,10) & 220.6174 & 61.0 & 92 & \\
CH$_3$CN & 12(5)--11(5) & 220.6410 & 247.3 & 40 & \\
-- & 12(4)-11(4) & 220.6792  & 183.1  & 54 & \\
-- & 12(3)-11(3) & 220.7090  & 133.1  & 150 &\\
-- & 12(2)-11(2) & 220.7302  & 97.4 & 96 & \\
-- & 12(1)-11(1) & 220.7430  & 76.0  & 295 & (1) \\
-- & 12(0)-11(0) & 220.7472  & 68.8  & 295 & (1) \\
$^{33}$SO$_2$ & 14(3,11)--14(2,12) & 220.9857 & 120.2 & 47 & \\
$^{34}$SO$_2$ & 22(2,20)--22(1,21) & 221.1149 & 248.1 & 116 & \\
-- & 13(2,12)--13(1,13) & 221.7357 & 92.5 & 153 & \\
SO$_2$ & 11(1,11)--10(0,10) & 221.9652 & 60.3 & 1397 & \\
H & $38\beta$ & 222.0117 & -- & 159 &(2) \\
CH$_3$CCH & 13(4)--12(4) & 222.0991 & 190.2 & 33 & (3) \\
-- & 13(3)--12(3) &  222.1288 & 139.6 & 118 & \\
CH$_3$OCHO $v=1$ & 18(6,12)--17(6,11) $A$ & 222.1488 & 312.3 & 106 & \\
CH$_3$CCH & 13(1)--12(1) & 222.1627 & 81.7 & 427 & \\
H$_2$CCO & 11(0,11)--10(0,10) & 222.1976 & 63.9 & 13 & (1) \\
-- & 11(3,8)--10(3, 7) & 222.2002 & 181.3 & 13 & (1) \\
\cline{1-6}
\multicolumn{6}{c}{USB Low} \\ 
\cline{1-6}
CO & 2--1 & 230.5380  & 16.5 & -- & \\
OCS & 19--18 & 231.0609  & 110.8 & 255 &  \\
$^{13}$CS & 5--4 & 231.2207  & 33.3 & 202 & \\
H & $30\alpha$ & 231.9009 & -- & 747 &\\
CH$_3$OCH$_3$ & 13(0,13)--12(1,12) $AA$ & 231.9877 & 80.9 & 122 & (1) \\
CH$_3$OCH$_3$ & 13(0,13)--12(1,12) $EA+AE$ & 231.9879 & 80.9 & 122 & (1) \\
\cline{1-6}
\multicolumn{6}{c}{USB High} \\ 
\cline{1-6}
CH$_3$OH  & 10(2,8)--9(3,7) $++$ &  232.4185 & 165.4 & 90 & \\
--  & $10(-3,8)-11(-2,10)$ & 232.9458 & 190.3 & 60 & (4) \\
CH$_3$CH$_2$CN & 26(6,21)--25(6,20) & 233.2050 & 190.9 & 47 & (1) \\
-- & 26(6,20)--25(6,19) & 233.2073 & 190.9 & 47 & (1) \\
-- & 42(4,38)--41(5,37) & 233.2910 & 410.1 & 33 & (1) \\
$^{34}$SO$_2$ & 10(5,5)--11(4,8) & 233.2964 & 109.8 & 33 & (1) \\
CH$_3$OCHO & 26(21,5)--27(20,7) $E$ & 233.7284 & 499.2 & 75 & \\
SO$_2$	& 28(3,25)--28(2,26) & 234.1870 & 403.0 & 591 &
\enddata
\tablenotetext{a}{Molecular lines detected above an intensity $I_\nu>50$ mJy beam$^{-1}$ at line peak in the subcompact-array maps, 
averaged over an area equal to the $5\sigma$ contour of the continuum map shown in Figure \ref{fig:BolocamSMAPB}, bottom.  
The first column refers to the molecule tag, the second column to the transition, 
the third to its rest frequency as found in Splatalogue (http://www.splatalogue.net/), the fourth 
to the upper-level energy, the fifth to the velocity-integrated flux in the all-configuration maps, and the fifth to additional notes.  
Data used from Splatalogue are compiled from the CDMS catalog \citep{Muller05} and the NIST catalog \citep{Lovas04}. 
A single flux is listed for blended lines. The $^{13}$CO and C$^{18}$O 2--1 maps 
have been combined with IRAM30m single-dish data.}
\tablenotetext{1}{Blended lines.}
\tablenotetext{2}{From the measured line intensity of the H$30\alpha$ line, and assuming LTE conditions, 
this feature is dominated by H$38\beta$. It 
could have a contribution of $<25\%$ from CH$_3$CCH 13(6)--12(6), $\nu_0=222.0144$ GHz.}
\tablenotetext{3}{From the measured line intensity of the H$30\alpha$ line, and assuming LTE conditions 
and a $10\%$ Helium abundance, this feature is dominated by CH$_3$CCH 13(4)--12(4). It could have 
a contribution of $<20\%$ from He$38\beta$,  $\nu_0=222.1022$ GHz.}
\tablenotetext{4}{Could be contaminated with CH$_3$CHO 4(4,1)--5(3,2) A--, $\nu_0=232.9518$ GHz.}
\end{deluxetable}
\end{center}

\end{document}